\documentclass[aps,prb,twocolumn,showpacs,amsmath,amssymb,superscriptaddress,floatfix]{revtex4-1}
\usepackage{graphicx}
\usepackage{dcolumn} 
\usepackage{bm} 
\usepackage{amsfonts}
\usepackage{amsmath}
\usepackage{ulem}
\usepackage{color}
\usepackage{hyperref}

\newcommand{\fab}[1]{}
\begin{document}

\title{Magnetic field effects on a nanowire with inhomogeneous Rashba spin-orbit coupling: Spin properties at equilibrium} 

\author{Fabrizio Dolcini}
\email{fabrizio.dolcini@polito.it}
\affiliation{Dipartimento di Scienza Applicata e Tecnologia del Politecnico di Torino, I-10129 Torino, Italy}

\author{Fausto Rossi}
\affiliation{Dipartimento di Scienza Applicata e Tecnologia del Politecnico di Torino, I-10129 Torino, Italy}

\begin{abstract} 
By modeling a Rashba nanowire contacted to leads via an inhomogeneous spin-orbit coupling profile, we investigate the equilibrium properties of the spin sector when a uniform magnetic field is applied along the nanowire axis. We find that the interplay between magnetic field and Rashba coupling generates a  spin current, polarised perpendicularly to the applied field and flowing through the nanowire even at equilibrium. In the nanowire bulk such effect persists far beyond the  regime where the nanowire mimics the helical states of a quantum spin Hall system, while in the leads the spin current is suppressed. Furthermore, despite the nanowire   not being proximized by superconductors,  at the  interfaces with the leads we predict the appearance of localized spin torques and spin polarizations, orthogonal to the magnetic field and partially penetrating into the leads. This feature, due to the inhomogeneity of the Rashba coupling, suggests to use caution   in interpreting spin polarization as signatures of Majorana fermions. 
When the magnetic field has a component also along the Rashba field, its collinearity with the spin polarization and  orthogonality to the spin current are violated   in the nanowire bulk too.
We analyze  these quantities in terms of the magnetic field and chemical potential for both long and short nanowires 
in experimentally realistic regimes.
\end{abstract}

\pacs{71.70.Ej, 78.67.Uh, 81.07.Gf, 85,75.-d}

\maketitle

\section{Introduction}

Rashba nanowires, i.e. semiconductor nanowires characterized by a strong Rashba spin-orbit coupling  (RSOC),  such as  InSb or InAs, are currently on the spotlight of a broad and growing scientific community, as they turn out to play a relevant role in various fields.
In spintronics, for instance, RSOC enables one to act electrically on the electron spin degree of freedom, with the fascinating perspective  to encode and manipulate information~[\onlinecite{nitta_2015}]. Furthermore it has been realized that, for sufficiently strong RSOC, a nanowire exposed to a magnetic field can effectively mimic the helical edge states of a quantum spin Hall (QSH)  system [\onlinecite{kane-mele2005a,kane-mele2005b,bernevig_science_2006}]  and that, when a superconducting film is further deposited on it, the proximized nanowire can realize a topological superconductor~[\onlinecite{vonoppen_2010},\onlinecite{dassarma_2010}].
In view of all these applications, a remarkable effort has been devoted in recent theoretical and experimental studies  to improve the tunability of the RSOC, reaching unprecedented high values of such coupling constant~[\onlinecite{gao-2012,wimmer_2015,nygaard_2016,sasaki_2017,loss_2017}].
Not only the bulk properties of   nanowires are interesting. The recent discovery that, under suitable conditions, Majorana fermions can be localized  at the interfaces  between a nanowire and a superconductor~[\onlinecite{vonoppen_2010},\onlinecite{dassarma_2010}] has been confirmed in a number of experiments~[\onlinecite{kouwenhoven_2012,liu_2012,heiblum_2012,xu_2012,defranceschi_2014,marcus_2016,marcus_science_2016}] and has provided a  major boost to the investigation of Rashba nanowires.

\begin{figure}[b]
\centering
\includegraphics[width=6.2cm,clip]{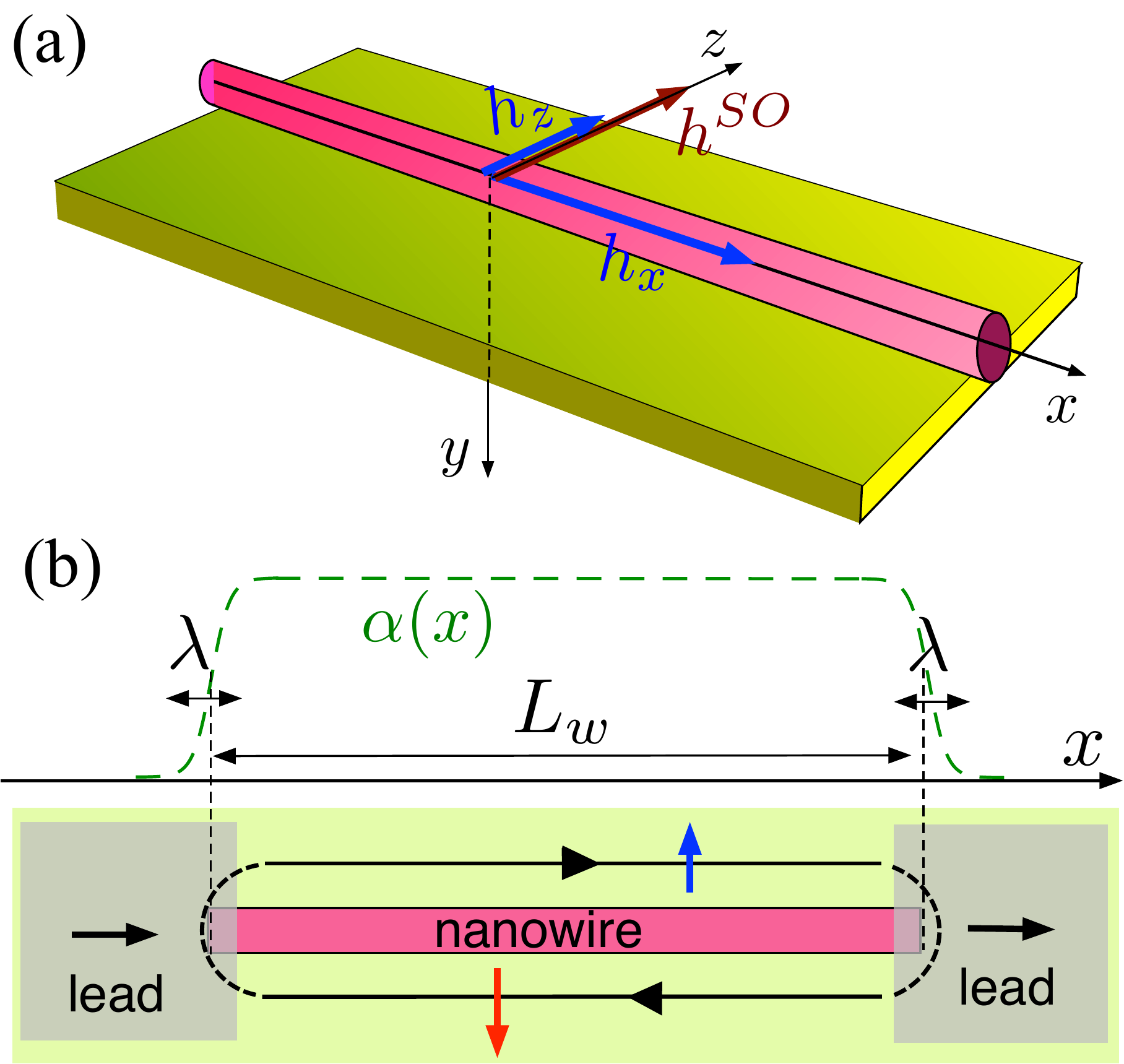}
\caption{(Color online) (a) A Rashba nanowire deposited on a substrate: the Rashba effective magnetic field $h^{SO}$ is directed along $z$, whereas an actual magnetic field,  externally applied  in the substrate plane, has  components $h_x$ (along the nanowire axis) and $h_z$ (along the  Rashba effective magnetic field).
(b) Top view of the Rashba nanowire contacted to leads: the setup can be modelled by an inhomogeneous Rashba coupling $\alpha(x)$  (dashed green curve) that varies from a finite value in the bulk of the nanowire   to zero in the leads, over a smoothing length $\lambda$.  Due to the inhomogeneous Rashba coupling, the leads act as magnetic barriers: spin flip processes  caused by the spin torque at the interfaces (red areas), form spin bound states carrying a spin current in the nanowire.}
\label{Fig-setup}
\end{figure}

Despite the huge interest that nanowires are  receiving  nowadays, various questions still remain mostly unexplored, so that a gap of understanding still exists in comparison to the analogous two-dimensional (2D) systems with RSOC. In the first instance, most theoretical works related to hybrid structures involving nanowires assume a uniform RSOC throughout the system. Since  the origin of RSOC is the strong electric field caused by structural inversion asymmetry (SIA), metallic or superconducting films deposited on top of a nanowire locally alter the RSOC~[\onlinecite{tokatly_PRB_2017}]. In fact, top or lateral gates are precisely exploited to this purpose. Similarly, the nanowire is contacted to ordinary metallic electrodes, where RSOC typically vanishes. 
The  effects of the  inhomogeneities of the RSOC are thus crucial and may affect the behavior of a nanowire-based setup, possibly even  in terms of its topological properties. However, while inhomogeneous RSOC has been discussed in  2D  systems like semiconductor quantum wells and graphene~[\onlinecite{brataas_2007,bercioux_2012,raimondi_2014,raimondi_2017,sherman_2018}], only a few groups have addressed it in one-dimensional (1D) systems~[\onlinecite{sanchez_2006,sanchez_2008,sherman_2011,sherman_2013,sherman_2017,loss_EPJB_2015}].

Secondly, as far as equilibrium properties are concerned, nanowires have been much less analyzed than 2D electron gases (2DEGs) with RSOC. In 2D, for instance, RSOC can lead to a background spin current flow even at equilibrium[\onlinecite{rashba_2003}], a surprising effect that opened up an interesting conceptual debate about its observability, with various proposals by several groups ~[\onlinecite{governale_2003,balseiro_2005,sonin_PRB_2007,sonin_PRL_2007,wang_PRL_2007,wang_PRB_2008,tokatly_2008,sablikov_2008,liang_PLA_2008,medina_2010,sonin_2010,nakhmedov_2012,liu_2014,liang_PLA_2015}]. In 1D systems, however,  such effect is absent[\onlinecite{malshukov_2003},\onlinecite{wang_2006}], and this has probably contributed to convey the impression that the equilibrium properties of Rashba nanowires are trivial. 
In fact, most works on nanowires  have focused on out of equilibrium properties, concerning either the charge sector, such as the behavior of the linear conductance as a function of the gate voltage or magnetic field~[\onlinecite{sanchez_2006,sanchez_2008,ojanen_2012,rainis-loss_PRL_2014,rainis-loss_PRB_2014,aguado_2015}], or the spin sector, such as the spin current under  time-dependent perturbations~[\onlinecite{malshukov_2003,malshukov_2005,sherman_2013,loss_2016}].   
Notably, in many of these works the presence of an external magnetic field is crucial.

This paper is devoted to show that in a nanowire+leads setup  the  interplay between a uniformly applied  magnetic field and the inhomogeneous RSOC leads to interesting effects in the spin sector, even at equilibrium and at low temperatures. First, while a mere RSOC or a magnetic field, separately, cannot cause any equilibrium spin current, the latter does arise in the nanowire bulk when both are present and the magnetic field is applied along the nanowire axis. Such equilibrium spin current is polarised along the Rashba field direction, i.e. orthogonally to the direction of the applied magnetic field.  Notably, while the spin current decays in the leads,  it persists in the  bulk far beyond the regime  where the nanowire mimics the helical states of the QSH effect, and can be tuned by the magnetic field and the chemical potential. 
Secondly, at the nanowire/lead interfaces we predict the appearance of  localized spin torque and  spin polarization orthogonal to the magnetic field, whose penetration into the leads is discussed in various regimes of chemical potential. 
Notably, this effect is qualitatively similar to the orthogonal spin polarization of the Majorana states localized at the boundaries of a proximized nanowire in the topological phase[\onlinecite{bena-simon_2012}], despite that in our case  there is no superconducting coupling and the nanowire is in the topologically trivial phase. 
This result, purely due to the inhomogeneity of the RSOC, suggests that caution should be taken in interpreting a localized and orthogonal spin polarization as a signature of Majorana fermions.
Furthermore, our analysis also shows that, in the inhomogeneous nanowire+leads system, the leads effectively act as magnetic barriers, and the interface spin torques as sources and sinks of the equilibrium spin current carried by spin bound states  in the nanowire, similarly to the charge current carried by the Andreev bound states in a superconductor/normal/superconductor junction. 
Finally, we address the effects of a magnetic field component parallel to the Rashba spin direction, showing that it leads to a spin polarization not collinear with the magnetic field and to a spin-orbit torque  also in the nanowire bulk. 

The paper is organized as follows. In Sec.~\ref{sec-2} we introduce the model and the quantities we shall analyze. Then, in Sec.~\ref{sec-3} we  discuss the bulk properties of the nanowire by taking the limit of  homogeneous RSOC. Specifically, we examine in details the origin of the emerging equilibrium spin current and analyze its behavior as a function of the applied magnetic field and chemical potential.    In Sec.~\ref{sec-4} we account for the presence of the leads by an inhomogeneous profile smoothly varying at the interfaces, and discuss the appearance of the spin torque and the orthogonal spin polarization at the  interfaces, as well as their penetration into the leads.   Section~\ref{sec-5} is devoted to analyze the effects of a magnetic field component parallel to the Rashba field direction. Finally in Sec.~\ref{sec-6}, after summarizing our main results,
 we conclude with a discussion of some possible experimental realizations and an outlook of future developments of this work.

\section{Model and spin equation of motion}

\label{sec-2} 
\subsection{System Hamiltonian}
We  consider a nanowire deposited on a substrate and  assume that one single electronic channel is active in the nanowire. The SIA emerging at the interface with the substrate gives rise to an effective Rashba spin-orbit ``magnetic" field ~$\mathbf{h}^{SO}$ orthogonal to the nanowire axis, in the substrate plane. We also consider the presence of an actual external magnetic field, applied along an arbitrary direction in the substrate plane.
For definiteness, we shall denote the longitudinal direction of the nanowire  by $x$, the direction perpendicular to the substrate plane and pointing downwards by $y$, while the direction of the Rashba field by $z$ [see Fig.\ref{Fig-setup}(a)].

Denoting by $\hat{\Psi}(x)= \left(   \psi_\uparrow(x) \,,\,\psi_\downarrow(x) \right)^T$ the electron spinor field, where $\uparrow,\downarrow$ correspond to spin projections along positive and negative Rashba field direction $z$, respectively, the Hamiltonian for a Rashba nanowire reads
\begin{equation}\label{H}
\hat{\mathcal{H}}  = \int \hat{\Psi}^\dagger(x)\,H(x)\,  \hat{\Psi}(x)\,dx
\end{equation}
where
\begin{equation}\label{H-realspace}
H(x)=  \frac{p_x^2}{2 m^*} \sigma_0 -\frac{\left\{ \alpha(x) , p_x\right\}}{2\hbar}  \sigma_z\, \, - \mathbf{h}\cdot \boldsymbol{\sigma} \quad,
\end{equation}
$p_x=-i\hbar \partial_x$ is the momentum operator, 
$\sigma_0$ the $2 \times 2$ identity matrix, and $\boldsymbol{\sigma}=(\sigma_x,\sigma_y,\sigma_z)$ are the Pauli matrices. Furthermore, 
 $\alpha(x)$ denotes the RSOC profile, in general inhomogeneous along the nanowire, so that the anticommutator with $p_x$ is necessary in Eq.(\ref{H-realspace}). Finally $\mathbf{h}=g \mu_B \mathbf{B}/2$ is the Zeeman energy vector induced by the external magnetic field~$\mathbf{B}=(B_x,0,B_z)$, with $\mu_B$ denoting the Bohr magneton and $g$ the Land\'e  factor. 
It is useful to decompose the Zeeman energy vector as $\mathbf{h}={h_x} \mathbf{i}_x+{h_z} \mathbf{i}_z$, where $h_x$ and $h_z$ denote the components parallel and perpendicular to the nanowire axis $x$, i.e. perpendicular and parallel to the Rashba spin-orbit field direction $z$, respectively  [see Fig.\ref{Fig-setup}(a)].   \\

Before focusing on the behavior of quantities in the spin sector, it is worth recalling an aspect related to the charge sector in systems with RSOC: While the charge density
has the usual expression
\begin{eqnarray}
\hat{n} &=&  {\rm e}\,\hat{\Psi}^\dagger(x) \,   \hat{\Psi}^{}(x) \quad,\label{charge-density-def} 
\end{eqnarray}
where ${\rm e}$ denotes the electron charge, the charge current density 
\begin{eqnarray}
\hat{J}^c\! &=& -\frac{i{\rm e}\hbar}{2 m^*} \left( \hat{\Psi}^\dagger(x) \, \partial_x\hat{\Psi}^{}(x) -\partial_x \hat{\Psi}^\dagger(x) \,  \hat{\Psi}^{}(x) \right) \,\,  \nonumber \\
& &  \hspace{1cm} -{\rm e}\frac{\alpha(x)}{\hbar} \sigma_z \hat{\Psi}^\dagger(x) \,   \hat{\Psi}^{}(x)  \label{charge-current-def}
\end{eqnarray}
includes a term [second line of Eq.(\ref{charge-current-def})] associated to the Rashba coupling $\alpha$ in the Hamiltonian (\ref{H-realspace}). Such term  originates from the fact that, in the presence of RSOC, the (charge) velocity operator becomes spin-dependent,  
\begin{equation}\label{vel-op-def}
v \doteq \frac{\left[ x, H(x)\right]}{i \hbar}=\frac{p_x}{m^*} -\frac{\alpha}{\hbar}\sigma_z \quad,
\end{equation}
and is essential to ensure the fulfilment of the charge continuity equation,
\begin{equation}\label{charge-cont-eq}
\partial_t \hat{n}+\partial_x \hat{J}^c=0\quad.
\end{equation}
In the following, we shall show that an analogous term plays a crucial role in the spin sector.

\subsection{Spin density, spin current, torques and spin equation of motion}
Let us now focus on the spin sector. We define the spin density  and spin current density operators in the standard way~as~[\onlinecite{rashba_2003},\onlinecite{sonin_2010}]
\begin{eqnarray}
\hat{\mathbf{S}} &=&\frac{\hbar}{2} \,\hat{\Psi}^\dagger(x) \, \boldsymbol\sigma\, \hat{\Psi}^{}(x) \label{spin-density-def} \\
\hat{\mathbf{J}}^s\! &=& \frac{1}{2} \left(  \hat{\Psi}^\dagger(x)\, \hat{\mathbf{S}} \, v \hat{\Psi}^{}(x) \, + {\rm H.c.} \right) =\nonumber \\
&=& \frac{\hbar}{2} \,\left(-\frac{i\hbar}{2 m^*} \left( \hat{\Psi}^\dagger(x) \, \boldsymbol\sigma\, \partial_x\hat{\Psi}^{}(x) -\partial_x \hat{\Psi}^\dagger(x) \, \boldsymbol\sigma\, \hat{\Psi}^{}(x) \right) \,\, \right.\nonumber \\
& & \left. \hspace{1cm} -\frac{\alpha(x)}{\hbar} \hat{\Psi}^\dagger(x) \, \frac{\{ \boldsymbol\sigma, \sigma_z\}}{2} \, \hat{\Psi}^{}(x) \right) \quad, \label{spin-current-def}
\end{eqnarray}
respectively. Formally, Eqs.(\ref{spin-density-def}) and (\ref{spin-current-def}) can be obtained from the related charge operators (\ref{charge-density-def}) and (\ref{charge-current-def}) by replacing ${\rm e} \rightarrow \hbar/2$ and by inserting the set $\boldsymbol{\sigma}$ of Pauli matrices between $\hat{\Psi}^\dagger $ and $\hat{\Psi}$, implying that electrons with opposite spins contribute with opposite sign to spin density and current, as compared to  the related charge   quantities. Note that, similarly to the second line of Eq.(\ref{charge-current-def}), the last term of   Eq.(\ref{spin-current-def}) stems from Rashba coupling, which modifies the spin velocity  by an extra term  along the $z$-direction of the Rashba field~$\mathbf{h}^{SO}$,
\begin{equation}\label{spin-vel-op-def}
\mathbf{v}^s \doteq \frac{\left\{ \boldsymbol\sigma, v  \right\}}{2}=\left( \frac{p_x}{m^*} \sigma_x \,,  \frac{p_x}{m^*} \sigma_y, \frac{p_x}{m^*} \sigma_z -\frac{\alpha}{\hbar} \right) \quad.
\end{equation}

Differently from charge, however, spin does not obey in general a continuity equation. For the Rashba nanowire one can prove that
\begin{equation}\label{spin-cont-eq}
\partial_t \hat{\mathbf{S}} +\partial_x \hat{\mathbf{J}}^s=\, \hat{\mathbf{T}}^{h}+\, \hat{\mathbf{T}}^{SO}  
\end{equation}
where  torque operators appear on the right-hand side [see Eq.(\ref{charge-cont-eq}) for comparison]. In particular, $\hat{\mathbf{T}}^{h}$ denotes the customary spin torque due to the magnetic field $\mathbf{h}$, 
\begin{equation}\label{Th-def}
\hat{\mathbf{T}}^{h} \doteq     \hat{\Psi}^\dagger       \left(   \boldsymbol\sigma \,  \times \mathbf{h}\right) \hat{\Psi}\quad,
\end{equation}
while $\hat{\mathbf{T}}^{SO}$ is an additional  spin-orbit torque  appearing in systems with RSOC, which
 can be given two  equivalent expressions
\begin{eqnarray}\label{TSO-def}
\hat{\mathbf{T}}^{SO} &\doteq&   \displaystyle \frac{\alpha(x)}{4} \left( \hat{\Psi}^\dagger \left[\boldsymbol\sigma, \sigma_z   \right]  \partial_x \hat{\Psi} -  \partial_x \hat{\Psi}^\dagger \left[\boldsymbol\sigma, \sigma_z   \right]  \hat{\Psi}  \right)\\
 & =& \displaystyle \frac{1}{2} \left( \hat{\Psi}^\dagger  ( \boldsymbol\sigma \,  \times  \mathbf{h}^{SO} )   \hat{\Psi}^{}+{\rm H.c.}  \right)\,   \quad,   
\end{eqnarray}
with
\begin{equation}\label{SO-field-op}
\mathbf{h}^{SO}(x,t)=\frac{\left\{\alpha(x),p_x\right\}}{2\hbar}(0,0,1)
\end{equation}
denoting the spin-orbit field operator (it actually has the dimension of an energy density). Note that, by definition, the spin-orbit torque $\hat{\mathbf{T}}_{SO}$ has components only in the plane $(x,y)$ orthogonal to the Rashba direction $z$. The proof of the equation of motion (\ref{spin-cont-eq}) is provided in the Appendix. \\

\subsection{Equilibrium expectation values}
The equilibrium properties of the system are obtained~as
\begin{equation}\label{equil-exp-val}
\begin{array}{rcl}
n(x) &=&\langle \hat{n} \rangle_\circ \\
J^c(x)&=&\langle \hat{J}^c \rangle_\circ \\
\mathbf{S}(x)&=&\langle \hat{\mathbf{S}} \rangle_\circ \\
\mathbf{J}^s(x)&=&\langle \hat{\mathbf{J}}^s \rangle_\circ \\
{\rm\mathbf{T}}^{h/SO}(x)&=&\langle \hat{\mathbf{T}}^{h/SO} \rangle_\circ  \quad,
\end{array}
\end{equation}
where $\langle \ldots \rangle_\circ$ denotes the quantum and statistical expectation value  over the equilibrium state. Note that, at equilibrium, these expectation values are time-independent. However,  they can be space-dependent in the case of   inhomogeneous RSOC, as we shall see below.

Besides the above quantities, it is also useful to introduce the particle density
\begin{equation}
\rho(x) \doteq \frac{n(x)}{\rm e}
\end{equation}
and the spin polarization, 
\begin{equation}\label{P-def}
\mathbf{P}(x) \doteq \frac{\langle \Psi^\dagger \boldsymbol\sigma \Psi^{} \rangle_\circ}{\langle \Psi^\dagger  \Psi^{} \rangle_\circ}= \frac{2}{\hbar} \,\frac{\mathbf{S}(x)}{\rho(x)}  \hspace{1.5cm} |\mathbf{P}|\le1 \quad.
\end{equation}
The dimensionless quantity $\mathbf{P}$ identifies, up to a universal constant, the spin density~$\mathbf{S}$ per electron, and is more straightforwardly  interpreted and customarily probed in experiments.

\section{The limit of homogeneous Rashba coupling:  nanowire bulk}
\label{sec-3}
\fab{\tiny NOTA INTERNA: Per motivi di convenienza nella presentazione dei risultati, qui nell'articolo ho deciso di semplificare la situazione rispetto al caso generale delle note. Ho assunto che il campo magnetico giaccia nel piano x-z (piano del sostrato). Questo corrispondere ad assumere che l'angolo $\phi_h$ delle note possa assumere solo i due valori $\phi_h=0$ (per componente perpendicolare diretta lungo $x$ positivi), oppure $\phi_h=\pi$, (per componente perpendicolare diretta diretta lungo $x$ negativi). Di fatto, $\phi_h$ non e' piu' necessario e conviene passare dalle coordinate sferiche alle coordinate circolari nel piano x-z. Quindi, mentre nelle note c'era ${h_x}>0$, $\theta_k \in  [0,\pi]$, e $\phi_h \in [0,2\pi]$, qui ho ${h_x}$ sia positivo che negativo, ho eliminato $\phi_h$ e sono passato ad un $\theta_k(here) \in  [-\pi,\pi]$. I calcoli delle note con ${h_x}>0$, $\theta_k \in  [0,\pi]$ e $\phi_h=0$ corrispondono al caso ${h_x}(here)>0$, $\theta_k(here)  \in  [0,\pi]$, mentre  i calcoli delle note con $\theta_k \in  [0,\pi]$ e $\phi_h=\pi$ corrispondono al caso ${h_x}(here)<0$,  $\theta_k \in  [-\pi,0]$. }

We start by analyzing the bulk properties of the nanowire~[\onlinecite{nota-bulk}], which can be addressed in the limit of very long nanowire length  $L_w \rightarrow \infty$ [see Fig.\ref{Fig-setup}(b)],  i.e. assuming a homogeneous Rashba coupling $\alpha(x) \equiv \alpha$. The Hamiltonian (\ref{H-realspace}) of the nanowire then commutes with~$p_x$. By expressing the electron spinor field in terms of its Fourier components
\begin{equation}\label{Psi-Fourier}
\hat{\Psi}(x)= \frac{1}{\sqrt{L_w}} \sum_k e^{i k x} \hat{\mathcal{C}}_k	\quad,
\end{equation}
with 
$\hat{\mathcal{C}}_k= \left(  \hat{c}_{k\uparrow} \,,\, \hat{c}_{k\downarrow}\right)^T$ denoting the Fourier mode operators,  the Hamiltonian  is block-diagonal in $k$-space, $\hat{\mathcal{H}}  = \sum_k  \hat{\mathcal{C}}^\dagger_k \, H_k^{}\,  \hat{\mathcal{C}}^{}_k   
$, where $H_k$ can be compactly written as
\begin{eqnarray}
H^{}_k &=&  \varepsilon^0_k \sigma_0 -(\alpha k+{h_z})\sigma_z-{h_x}  \sigma_x  \hspace{0.5cm} \label{Hk}\\
&=& \varepsilon^0_k \sigma_0 -\sqrt{(\alpha k +{h_z})^2+h_x^2}\, \, \mathbf{n} \cdot \boldsymbol{\sigma} \quad. \nonumber 
\end{eqnarray}
Here $\varepsilon^0_k=\hbar^2 k^2/2 m^*$, while the unit vector
\begin{equation}\label{n-vec-def}
\mathbf{n}(k) \doteq \left( \sin\theta_k \,,0 \,,\, \cos\theta_k \right)
\end{equation}
identifies the  $k$-state spin-orientation in the $x$-$z$ plane. Here the   angle~$\theta_k \in [-\pi;  \pi ]$ is defined through
\begin{equation}\label{thetak-def}
\left\{\begin{array}{lcl}
\cos \theta_k  &=&\displaystyle  \frac{\alpha k+{h_z}}{\sqrt{(\alpha k +{h_z})^2+h_x^2}}\\  
\sin \theta_k  &=&\displaystyle \frac{{h_x}}{\sqrt{(\alpha k +{h_z})^2+h_x^2}}
\end{array}\right.  
\end{equation}
and depends on both the magnetic field  and $k$, due to the RSOC $\alpha$. From the expression (\ref{Hk}), one  straightforwardly obtains the two spectrum bands  
\begin{equation}\label{spectrum}
E_\pm(k)=  \, \varepsilon^0_k\pm \sqrt{h_x^2+(\alpha k +{h_z})^2}
\end{equation}   
as well as the related eigenvectors
\begin{eqnarray}
\label{eigenvectors}
w_{k-}= \left(
\begin{array}{c}
\cos \frac{\theta_k}{2} \\  \\ 
 \sin \frac{\theta_k}{2}\,  
\end{array}\right)\hspace{0.5cm} 
w_{k +}= \left(
\begin{array}{c}
-  \sin \frac{\theta_k}{2}\\  \\ 
\cos \frac{\theta_k}{2}
\end{array}\right)\,. \hspace{0.5cm} 
\end{eqnarray}
Thus, by performing a rotation $U_k=\exp[-i \theta_k  \sigma_2/2]$ by the angle $\theta_k$, the Fourier mode operators $\hat{\mathcal{C}}_k$ can be re-expressed  in terms of new fermionic operators $\hat{\Gamma}_k=(\hat{\gamma}_{k-} ,\hat{\gamma}_{k+})^T$ related to the eigenvectors (\ref{eigenvectors}),  
\begin{equation}\label{C-Gamma}
\hat{\mathcal{C}}_k =U_k \, \hat{\Gamma}_k = w_{k-} \hat{\gamma}_{k-} + w_{k+} \hat{\gamma}_{k+}     \quad,
\end{equation}
and the Hamiltonian straightforwardly acquires a diagonal form
\begin{equation}
\hat{\mathcal{H}}=\, \, \sum_k \left(E_{-}^{}(k) \hat{\gamma}^\dagger_{k-} \hat{\gamma}^{}_{k-} \, + E_{+}^{}(k) \hat{\gamma}^\dagger_{k+} \hat{\gamma}^{}_{k+} \right)\quad.
\end{equation}
Notably, in the presence of a perpendicular component ${h_x}  \neq 0$ of the Zeeman field (i.e. $\theta_k \neq 0,\pi$),  the eigenstates $\Psi_{k\pm}(x)=w_{k\pm} e^{i k x}$ diagonalizing the Hamiltonian are {\it not} eigenstates of the (charge) velocity operator $v$, due to the Rashba spin-orbit term appearing in Eq.(\ref{vel-op-def}). This aspect will be important in interpreting the spin current, as we shall discuss below. 
Yet, the quantum expectation value of $v$ on   each eigenstate  corresponds of course to the group velocity associated to the slope of the spectrum,
\begin{eqnarray}
v_{\pm}(k) &\doteq &\langle \Psi_{k\pm}| v |\Psi_{k\pm} \rangle  = \frac{\hbar k}{m^*} \pm \frac{\alpha}{\hbar} \cos\theta_k = \nonumber \\
&=& \frac{1}{\hbar} \frac{\partial E_{\pm}}{\partial k}\quad,
\end{eqnarray}
where $E_\pm$ is given in Eq.(\ref{spectrum}).\\

Using Eqs.(\ref{Psi-Fourier}) and  (\ref{C-Gamma}) to re-express the  operators Eqs.(\ref{charge-density-def}), (\ref{charge-current-def}), (\ref{spin-density-def}), (\ref{spin-current-def}), (\ref{Th-def}) and (\ref{TSO-def}) in terms of the diagonalizing operators $\hat{\gamma}_{k\pm}$'s, and   exploiting the fact that, at equilibrium
\begin{equation}
\langle \hat{\gamma}^{\dagger}_{k \, b}\hat{\gamma}^{}_{k^\prime\,b^\prime} \rangle_\circ= \delta_{k,k^\prime} \delta_{b,b^\prime} f^\circ(E_{b}(k)) \hspace{1cm} b,b^\prime=\pm 
\end{equation}
where $f^\circ(E)=\{1+\exp[(E-\mu)/k_B T]\}^{-1}$ is the Fermi distribution  at temperature $T$ and chemical potential $\mu$, the equilibrium expectation values (\ref{equil-exp-val}) are straightforwardly obtained. 
In particular, in the bulk limit, $L_w \rightarrow \infty$, one can pass to the continuum. For  the charge sector one obtains
\begin{eqnarray}
n  &=& {\rm e}  \int \frac{dk}{2\pi} \left[\, f^\circ(E_{-}(k))+  \, f^\circ(E_{+}(k))\right] \label{ncharge-equil}  \\
J^c  &=&   {\rm e} \sum_{b=\pm}  \int \frac{dk}{2\pi}\left( \frac{\hbar k}{m^*}\, +b \, \frac{\alpha}{\hbar}  \, \cos\theta_k\, \right) \, f^\circ(E_{b}(k)) \,\,\,\label{Jc-homo}
\end{eqnarray}
and by combining Eq.(\ref{spectrum}) and (\ref{Jc-homo})  one   straightforwardly sees that $J^c={\rm e} (2\pi \hbar)^{-1} \sum_b \int dk\, \partial_k E_b(k) f^\circ(E_b(k))=0$, i.e.  the   charge current  vanishes, as expected at equilibrium.  In contrast, for the spin sector one finds
{\small
\begin{eqnarray}
S_{x} &=& \frac{\hbar}{2}\int \frac{dk}{2\pi}\sin\theta_k  \left[\, f^\circ(E_{-}(k))-  \, f^\circ(E_{+}(k))\right] \hspace{1cm} \label{Sx-gen}  \\
S_{y}  &=&  0 \hspace{0.5cm} \label{Sy-gen} \\
S_{z} &=& \frac{\hbar}{2}\int \frac{dk}{2\pi} \cos\theta_k  \left[\, f^\circ(E_{-}(k))-  \, f^\circ(E_{+}(k))\right] \label{spar-equil}  \hspace{1cm} \label{Sz-gen} 
\\
J^s_x &=& \frac{\hbar}{2} \int \frac{dk}{2\pi}  \,\frac{\hbar k}{m^*}\, \sin\theta_k \, \left[\, f^\circ(E_{-}(k))-  \, f^\circ(E_{+}(k))\right]  \hspace{1cm}  \label{Jsx-gen} \\
J^s_y &=& 0 \label{Jsy-gen} \\
J^s_z &=&  -\frac{\hbar}{2} \sum_{b=\pm} b \int \frac{dk}{2\pi}  \left(  \frac{\hbar k}{m^*}\, \cos\theta_k \,+b\frac{\alpha}{\hbar} \right)  \, f^\circ(E_{b}(k))  \hspace{1cm}
\label{Jsz-gen}
\\
{\rm T}^{h}_{x} &=& 0 \label{Thx-gen} \\
{\rm T}^{h}_{y} &=& \sum_{b=\pm} b \int \frac{dk}{2\pi}  ({h_z} \sin\theta_k-\,{h_x} \cos\theta_k)    f^\circ(E_{b}(k)) \label{Thy-gen} \\
{\rm T}^{h}_{z} &=& 0  \label{Thz-gen}
\\ & & \nonumber\\
{\rm T}^{SO}_x &=& 0 \label{TSOx-gen} \\
{\rm T}^{SO}_y &=& -\int \frac{dk}{2\pi}  \,\alpha k\, \sin\theta_k \, \left[\, f^\circ(E_{-}(k))-  \, f^\circ(E_{+}(k))\right]  \label{TSOy-gen} \\
{\rm T}^{SO}_z &=& 0\label{TSOz-gen} 
\end{eqnarray}
}
\noindent 
Note that, since we are considering at the moment the {\it homogeneous}  bulk of the nanowire, all   above equilibrium quantities are independent of the space coordinate $x$, besides being independent of time. Then,  the   expectation values  of the torque operators appearing on the left hand side of Eq.(\ref{spin-cont-eq}) vanish, 
\begin{equation}\label{torque-cancellation}
{\rm \mathbf{T}}^{h}+ {\rm \mathbf{T}}^{SO}\,=0
\end{equation}
and the spin continuity equation is fulfilled. This  can also be directly deduced by summing up Eq.(\ref{Thy-gen}) and Eq.(\ref{TSOy-gen})  and by using Eq.(\ref{thetak-def}).  

Here below we shall analyze the above quantities as a function of the magnetic field and chemical potential,   focusing on the spin sector. 

\begin{figure}[t]
\centering
\includegraphics[width=7.5cm]{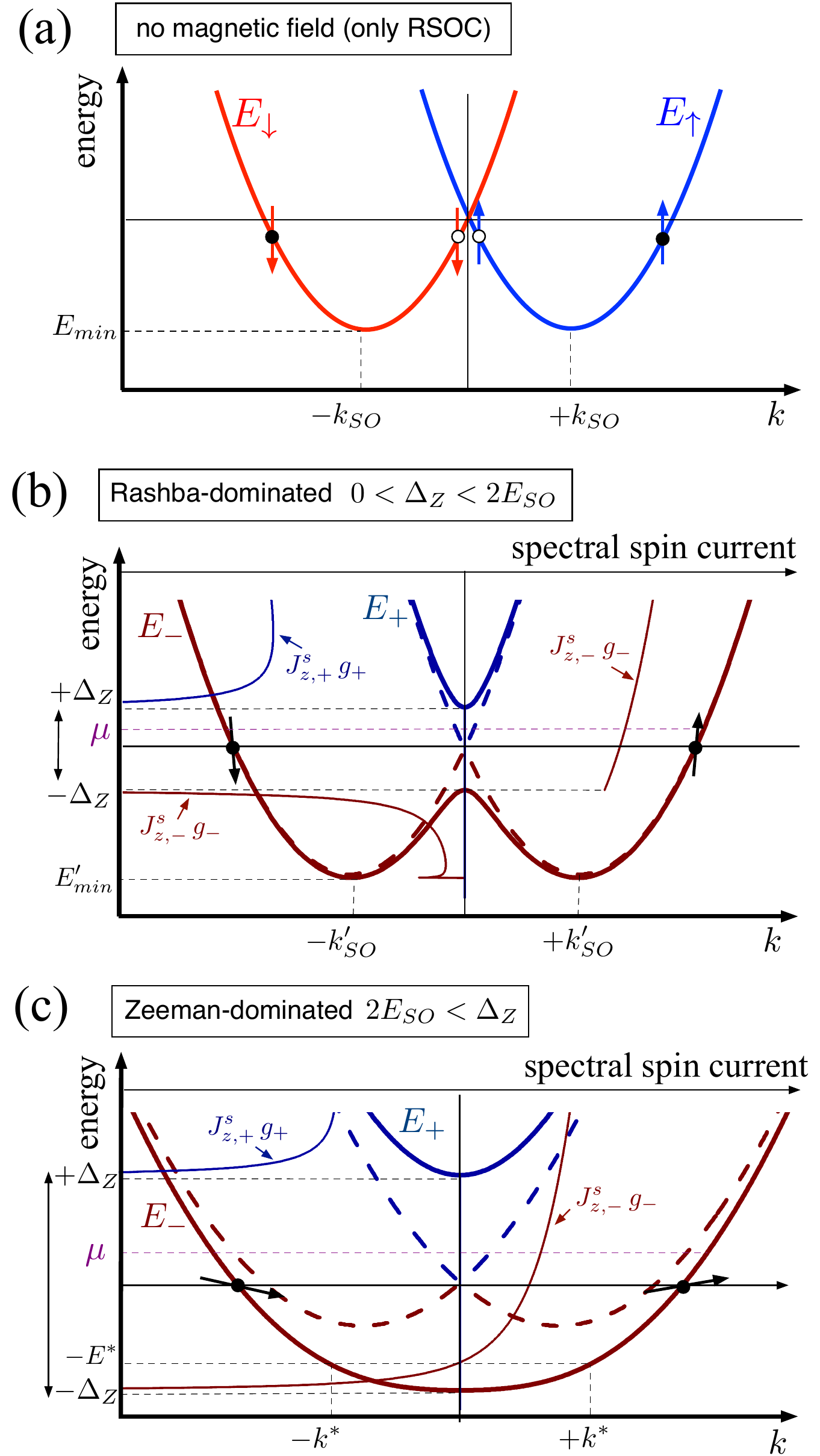}
\caption{(Color online) The bulk spectrum of the Rashba nanowire (homogeneous  RSOC).  (a) In the absence of magnetic field, spin is a good quantum number and the spectrum consists of  two bands $E_\uparrow$ and $E_\downarrow$. Vertical arrows describe the spin orientation  for the case $\alpha>0$.
(b) Effects of a weak magnetic field $h_x$: The Rashba-dominated regime. Thick  solid curves denote the electronic spectrum (left axis) in the presence of a magnetic field ${h_x}$ applied along the nanowire axis, i.e. perpendicularly to the Rashba spin-orbit direction $z$. Arrows denote the spin orientation when $\alpha>0$. Thick dashed curves denote the spectrum for $h_x=0$ and are a guide to the eye. Thin solid  curves  describe   the spectral weight $J^s_{z,\pm}(E)\, g_\pm(E)$ of the two bands [the reader should refer to the left vertical axis (energy) and to the upper horizontal axis (spectral spin current)]. (c) Effects of a strong magnetic field: in the Zeeman-dominated regime, the minimum of the bands is at $k=0$. The meaning of the curves is the same as in (b).  }
\label{Fig2}
\end{figure}

\subsection{Vanishing magnetic field}   
\label{sec-3-A}
Let us first briefly recall the effects of a purely Rashba coupling $\alpha$ on the spin sector. For vanishing magnetic field, ${h_x}={h_z}=0$, the Hamiltonian (\ref{Hk}) commutes with~$\sigma_z$, so that the spin quantization axis is $z$ for all electronic states of the nanowire, i.e. $\theta_k=0$ or $\pi$ in Eq.(\ref{thetak-def}). 
The Rashba coupling thus leads to the well known spectrum displayed by the solid curves of Fig.\ref{Fig2}(a):  the spin-$\uparrow$ and spin-$\downarrow$ states are horizontally displaced in opposite directions by the Rashba wavevector
\begin{equation}
\label{kSO-def}
k_{SO}=\frac{m^* |\alpha|}{\hbar^2}\quad,
\end{equation}
and the minima of the parabolic spectrum are lowered to $E_{min}=-E_{SO}$, where 
\begin{equation}
\label{ESO-def}
E_{SO}=\frac{m^*\alpha^2}{2 \hbar^2} =\frac{\hbar^2 k_{SO}^2}{2 m^*} 
\end{equation} 
is   the Rashba spin-orbit energy.
Inserting $\sin\theta_k =0$ and $\cos\theta_k=\mbox{sgn}(\alpha k)$ in the general expressions (\ref{Sx-gen})-(\ref{Thz-gen}) one can show that 
\begin{equation}\label{case-1}
\mathbf{P}=\mathbf{J}^s={\rm\mathbf{T}}^{h}={\rm\mathbf{T}}^{SO}=0\quad.
\end{equation}
The absence of a net spin polarization is due to the fact that the spin-$\uparrow$ and $\downarrow$ bands, despite being displaced, lead to perfectly opposite contributions at each energy $E$. The absence of   spin current can be understood from Fig.\ref{Fig2}(a) as follows: at any energy $E$, each pair of time-reversed states $(k\uparrow, -k\downarrow)$ is characterized by opposite group velocities and opposite spin orientations (filled circles), thereby carrying a  spin current. However, at the same energy $E$, another time-reversed pair $(k^\prime\downarrow,-k^\prime\uparrow)$ exists, with the same group velocities but reversed spin orientations  (empty circles), whose   spin current cancels the one of the former pair.
Finally, since the spin torque ${\rm\mathbf{T}}^{h}$ is trivially absent for vanishing magnetic field, Eq.(\ref{torque-cancellation}) implies that the spin-orbit torque ${\rm\mathbf{T}}^{SO}$ vanishes too. \\As summarized in Eq.(\ref{case-1}),
the mere presence of a RSOC has no effect on the nanowire spin sector at equilibrium.

\subsection{Effects of a perpendicular magnetic field: appearance of an equilibrium spin current}
\label{sec-perp-field}
When a magnetic field ${h_x}$ is applied along the nanowire axis $x$, i.e. perpendicularly to the spin-orbit direction~$z$, the scenario changes significantly. The spin orientation Eq.(\ref{n-vec-def}) of each electron  state becomes $k$-dependent [see Eq.(\ref{thetak-def})].
Furthermore, the two bands do not intersect anymore, and at $k=0$ they are separated by a gap $2 \Delta_Z$ in the spectrum (\ref{spectrum}), with
\begin{equation}\label{DeltaZ-def}
\Delta_Z=|h_x|=\left| \frac{g\, \mu_B}{2}  B_x \right| \quad.
\end{equation} 
Two regimes can be identified, as illustrated in Fig.\ref{Fig2}. For a weak applied field, namely in the Rashba-dominated regime $\Delta_Z< 2 E_{SO}$ [see Fig.\ref{Fig2}(b)],  the effect of the magnetic field is i) to {\it decrease}~to  the value
\begin{equation}\label{kprime-def}
k^\prime_{SO}\doteq k_{SO}\sqrt{1-\frac{\Delta_Z^2}{4 E_{SO}^2}}
\end{equation}  
the magnitude of the wavevectors $k=\pm k^\prime_{SO}$  corresponding to the minima of the lower band~$E_{-}$, and ii) to correspondingly  {\it lower} the value of such minima,   
\begin{equation}
\label{Eminprime-Rashba}
E^\prime_{min}=-E_{SO}\left(1+\frac{\Delta_Z^2}{4 E^2_{SO}}\right)\quad.
\end{equation}
In contrast, for stronger fields, i.e. in the Zeeman-dominated regime  $\Delta_Z> 2 E_{SO}$  [see Fig.\ref{Fig2}(c)],  the lower band has only one minimum   at $k=0$, with energy
\begin{equation}
E^\prime_{min}=-\Delta_Z\quad. \label{Eminprime-Zeeman}
\end{equation}
These essentially different features with respect to the case of vanishing magnetic field entail deviations from the trivial situation of the spin sector  Eq.(\ref{case-1}), and give rise to two effects. 
The first one is the expected emergence of a spin polarization along the direction of the applied magnetic field~${h_x}$
\begin{equation}
\mathbf{P}=(P_x,0,0) \label{P-case3}  
\end{equation}
with 
\begin{eqnarray}
P_{x} &=&  \frac{1}{\rho} \! \int \frac{dk}{2\pi} \,\frac{h_x  \,\left[ f^\circ(E_{-}(k)) -f^\circ(E_{+}(k))\right]}{\sqrt{(\alpha k)^2+h_x^2}}  \label{Px-ris}  \quad,
\end{eqnarray}
as can be deduced from Eq.(\ref{Sx-gen}), Eq.(\ref{Sz-gen}) and (\ref{P-def}).
An inspection of Eq.(\ref{Px-ris}) shows  that this effect is essentially due to the magnetic field, since it exists also without RSOC, i.e. for $\alpha=0$. However, the RSOC $\alpha \neq 0$ does affect the dependence of $P_x$ on~$h_x$. Indeed for each $k$-state the spin lies in the $x$-$z$ plane and forms an angle~ $\theta_k$ with the $z$-axis [see Eq.(\ref{n-vec-def})]. Thus, although  the spin $z$-component of  states with opposite $k$'s mutually cancel out and leave a net spin polarization  directed along~$x$, the latter is only a fraction $\sin\theta_k$ of the available polarization [see Eq.(\ref{thetak-def})]. 
\\

\begin{figure} 
\centering
\includegraphics[width=\columnwidth,clip]{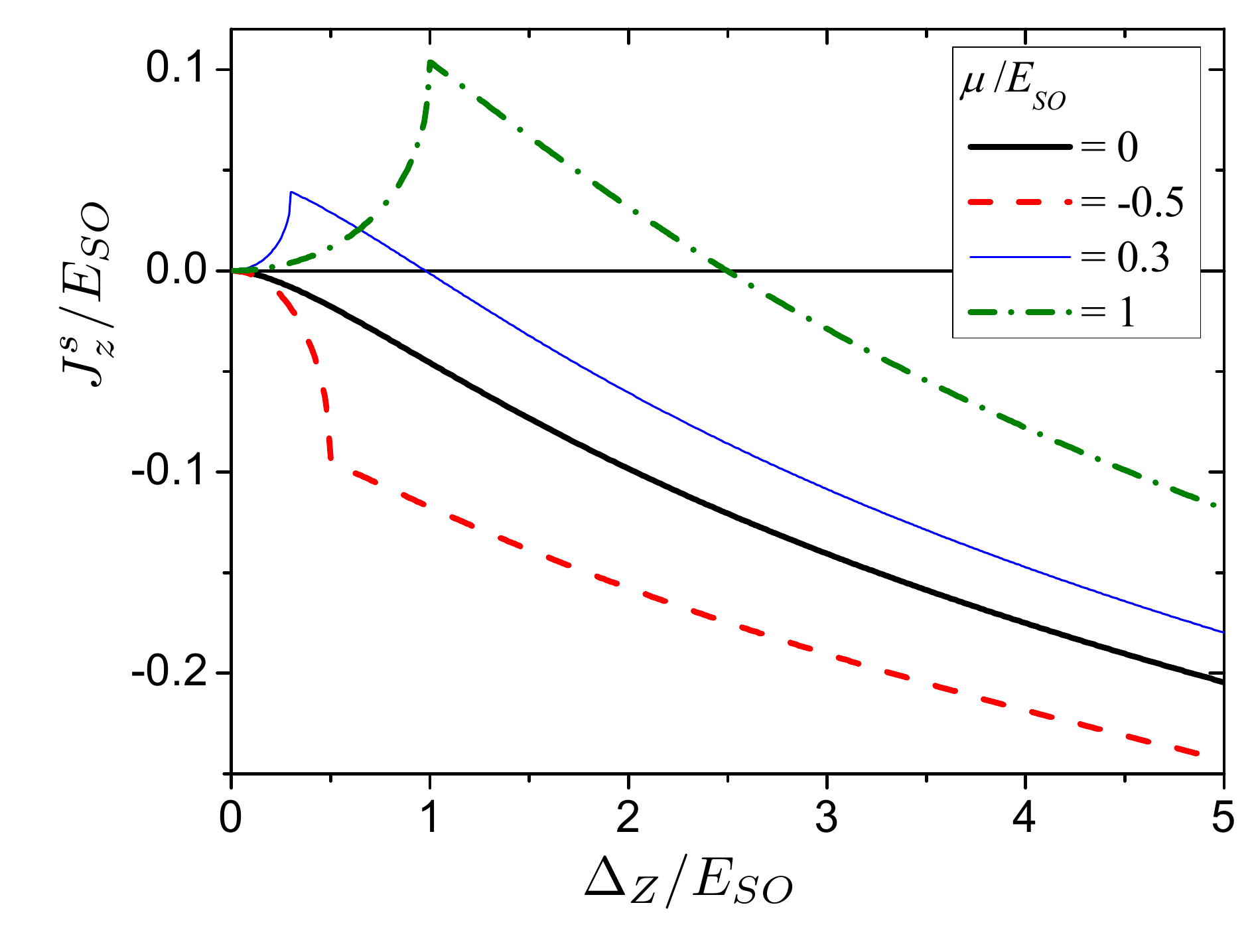}\\
\caption{(Color online)  The zero temperature behavior of the equilibrium spin current $J^s_z$, induced by the interplay between the applied magnetic field and the RSOC, is shown as a function of the ratio $\Delta_Z/E_{SO}$ of the Zeeman gap energy Eq.(\ref{DeltaZ-def}) to the Rashba spin-orbit energy Eq.(\ref{ESO-def}), for various values of the chemical potential~$\mu$.}
\label{Fig3-Jsz-hperp}
\end{figure}

The second effect is not trivial, and is one of our main results: It consists in the appearance of an equilibrium spin current  flowing through the nanowire and polarized {\it along} the Rashba direction $z$, i.e.  {\it perpendicularly} to direction $x$ of  the applied magnetic field ${h_x}$. Explicitly, from Eqs.(\ref{Jsx-gen})-(\ref{Jsy-gen}) and (\ref{Jsz-gen}), one finds 
\begin{equation}\label{Js-case3}
\mathbf{J}^s=(0,0,J^s_z)
\end{equation}
with
{\small
\begin{eqnarray}
J^s_z =  - \sum_{b=\pm}  \! \int \! \frac{dk}{2\pi}  \left( \frac{\alpha}{2}+b \frac{\hbar^2 k}{2m^*} \frac{\alpha k}{\sqrt{(\alpha k)^2+h_x^2}}  \right)  f^\circ(E_{b}(k)) ,  \quad \,\label{Jsz-hperp}
\end{eqnarray}
}

\noindent where $E_\pm(k)$ is given in Eq.(\ref{spectrum}). The behavior of  $J^s_z$ is illustrated in Fig.~\ref{Fig3-Jsz-hperp} as a function of the ratio $\Delta_Z/E_{SO}$ of the Zeeman gap energy to the Rashba spin-orbit energy, for various values of the chemical potential~$\mu$, in the case $\alpha>0$. As one can see, while for $\mu \le 0$ its magnitude monotonously increases with the magnetic field, for $\mu>0$ the behavior is non monotonous. Furthermore, cusps arise for any $\mu \neq 0$. 
The physical  origin of  the spin current and its behavior deserve a detailed analysis, which will be carried out in the next subsection. Here we just mention that,  although for simplicity of presentation we have assumed here a magnetic field applied along the nanowire axis $x$, the result (\ref{Jsz-hperp}) for the equilibrium spin current holds for any magnetic field $\mathbf{h}_\perp=(h_x,h_y,0)$ lying in the $(x,y)$ plane orthogonal to the Rashba direction $z$, upon the replacement $|h_x| \rightarrow |\mathbf{h}_\perp|$. Instead, the case of a magnetic field with a component along the Rashba direction $z$ will be analyzed in Sec.~\ref{sec-5}.
\\

We conclude this subsection by a comment related to the torques. In the present case, where the magnetic field $\mathbf{h}=(h_x,0,0)$ is applied along the nanowire axis $x$, the equilibrium spin density vector $\mathbf{S}$ and the polarization $\mathbf{P}$ are both collinear with $\mathbf{h}$ [see Eq.(\ref{P-case3})]. One can thus deduce from Eqs.(\ref{spin-density-def}) and (\ref{Th-def}) that the spin torque ${\rm \mathbf{T}}^{h}$ vanishes. 
Then, the spin-orbit torque ${\rm \mathbf{T}}^{SO}$ must vanish as well, because of Eq.(\ref{torque-cancellation}). In conclusion, for a magnetic field applied along the nanowire axis, both torques separately vanish
\begin{equation}\label{torques-case3}
{\rm \mathbf{T}}^{h}={\rm \mathbf{T}}^{SO}=0\quad.
\end{equation} 
This can also be deduced from Eqs.(\ref{TSOy-gen}) and (\ref{Thy-gen}) by noticing that, for ${h_z} =0$, $\sin\theta_k$ and $\cos\theta_k$ are even and odd in $k$, respectively [see Eqs.(\ref{thetak-def})], and all integrals therein   vanish by antisymmetry.

\subsection{Origin and meaning of the bulk equilibrium spin current}
Let us now discuss the physical origin and the behavior of the equilibrium spin current found in Eq.(\ref{Jsz-hperp}) and shown in Fig.\ref{Fig3-Jsz-hperp}. 
We start by  specifying the conditions for its appearance, which can be done analyzing some special limits of Eq.(\ref{Jsz-hperp}). On the one hand, when $\alpha=0$  the spin current vanishes.  On the other hand, when ${h_x}=0$  the band spectrum  reduces to the case of the purely Rashba coupling [see Fig.\ref{Fig2}(a)] where  the net spin current also vanishes, as observed in Sec.\ref{sec-3-A}. Thus, differently from the spin polarization (\ref{Px-ris}), the equilibrium spin current needs the presence of {\it both} the Rashba coupling $\alpha \neq 0$ and the perpendicular magnetic field ${h_x} \neq 0$.   We thus emphasize the difference with respect to the case of 2D systems with RSOC, where an equilibrium spin current has been predicted to arise already without magnetic field, i.e. when time-reversal symmetry is preserved[\onlinecite{rashba_2003}]. For nanowires the presence of the magnetic field  is crucial.

Secondly, we observe that the direction of flow of the spin current is determined by  the sign of the RSOC $\alpha$, and {\it not} by the direction of the magnetic field along the nanowire axis.   Indeed  from Eqs.(\ref{spectrum}) and (\ref{Jsz-hperp})  one can rewrite $J_z^s=(2\pi)^{-1}\int dk \sum_{b=\pm} J_{z,b}^s(k) f^\circ(E_b(k))$, where each spin current contribution 
\begin{equation}\label{Jsz-as-alpha-der} 
J_{z,b}^s(k)=-\frac{\hbar^2}{2 m^*} \frac{\partial \mathcal{E}_b(k)}{\partial \alpha} 
\end{equation}
can be regarded to as the response to the RSOC $\alpha$ of the energy $\mathcal{E}_b(k)=E_b(k)+E_{SO}$ measured from the band bottom $-E_{SO}$. In particular
 $J_z^s$
is an {\it odd} function of the Rashba coupling~$\alpha$  and an {\it even} function of the magnetic field $h_x$, i.e. it depends only on the gap energy $\Delta_Z$ [see Eq.(\ref{DeltaZ-def})].  
Notice the striking contrast with the  spin polarization in Eq.(\ref{P-case3}):  $\mathbf{P}$  is  pointing along $x$, i.e. orthogonally to the equilibrium spin current polarization Eq.(\ref{Js-case3}), it exists also for vanishing RSOC, is an odd function of the magnetic field $h_x$ and an even function of $\alpha$ [see  Eq.(\ref{Px-ris})]. Indeed each contribution to the spin polarization $\mathbf{P}$ can be expressed as the response of the energy $\mathcal{E}_b(k)$ to the applied magnetic field $\mathbf{h}$,"
\begin{eqnarray}
\mathbf{P}_b(k)=-\rho^{-1} \nabla_{\mathbf{h}}\mathcal{E}_b(k)\quad, \label{P-as-h-der} 
\end{eqnarray}
as straightforwardly obtained from Eqs.(\ref{spectrum}) and (\ref{Sx-gen})-(\ref{Sz-gen}).
 The  quantities $J_z^s$ and $\mathbf{P} $ are thus related only indirectly. In particular,  the spin polarization component $P_z$ parallel to the spin current polarization vanishes  [compare Eqs.(\ref{P-case3}) and (\ref{Js-case3})], implying that $J^s_z$ is not due to an unbalance of the number of spin-$\uparrow$ carriers with respect to spin-$\downarrow$ carriers.

In order to understand the origin of the spin current  induced by the magnetic field at  equilibrium, we recall its physical meaning, which can be straightforwardly read off from its definition [see first line of Eq.(\ref{spin-current-def})]: the spin current consists, up to the symmetrization of the non-commuting operators $\hat{\mathbf{S}}$ and $v$, in a non-vanishing expectation value  $\langle \hat{\mathbf{S}} v\rangle$. 
 Such expectation value can always be rewritten as the sum of two terms[\onlinecite{sonin_2010}]. The first one is related to the  product $\langle \hat{\mathbf{S}}\rangle\langle  v\rangle$, and describes the fact that a charge current transfers spin, when a  spin unbalance occurs along some orientation. The second one encodes the  correlation between between  velocity $v$ and  spin $\hat{\mathbf{S}}$. When the average velocity $\langle  v\rangle$ vanishes, the latter term is the only surviving contribution, and is also referred to as the {\it pure} spin current~[\onlinecite{sonin_2010}]. This is precisely what we find in the Rashba nanowire at equilibrium when the magnetic field is applied. As the effect survives down to zero temperature, the   spin current Eq.(\ref{Jsz-hperp})  describes the {\it non-trivial} quantum correlation  between the propagation direction $v$ and the spin orientation  $\hat{\mathbf{S}}$. The ground state    is thus characterized by electrons with opposite spins propagating in opposite directions,  corresponding to a pure transport of spin, without net charge current.   

As far as  1D systems are concerned,  this phenomenon is known to occur in Rashba nanorings[\onlinecite{governale_2003},\onlinecite{wang_PRL_2007},\onlinecite{medina_2010}], where the circular geometry induces a spin-texture of the RSOC, leading to a spin Berry phase[\onlinecite{loss-goldbart_PRL_1990},\onlinecite{loss-goldbart_PRB_1992},\onlinecite{aronov_1993}]: Electrons with opposite spin polarizations circulate clockwise and counter-clockwise along the ring, carrying a persistent spin current without an accompanying charge current.
An equilibrium spin current also arises in the 1D helical edge states of a QSH system, where right-moving electrons are characterized by (say) spin-$\uparrow$ and left-moving electrons by spin-$\downarrow$. At equilibrium, there is an equal number of right- and left-movers, so that there is no net velocity in the system ($\rightarrow \langle v\rangle=0$), no net spin polarization ($\rightarrow\, \langle \hat{S}_z \rangle=0$), but a   spin current  ($\rightarrow \langle   \hat{S}_z\, v\rangle \neq 0$), corresponding to a spin flow from left to right without charge flow. This effect is a straightforward consequence of the velocity-spin locking characterizing the helical QSH edge states~[\onlinecite{hugues_2014}].

In single-channel Rashba nanowires, where the RSOC has no spin texture, there is no bulk equilibrium spin current without magnetic field[\onlinecite{malshukov_2003},\onlinecite{wang_2006}]. However, as shown above, it does appear  when a uniform magnetic field $h_x$ is added.
Notably, for weak magnetic fields (Rashba-dominated regime) 
the states inside the magnetic gap of the nanowire are known to effectively behave like QSH edge states~[\onlinecite{vonoppen_2010,dassarma_2010,alicea_PRB_2010,kouwenhoven_2012,liu_2012,heiblum_2012,xu_2012,defranceschi_2014,marcus_2016,marcus_science_2016}], where the equilibrium spin current is known to flow [\onlinecite{hugues_2014}].  One is thus tempted to argue that this analogy with QSH edge states also explains why the equilibrium spin current arises. 
We anticipate that this argument is not sufficient to account for the features of the equilibrium spin current shown in Fig.\ref{Fig3-Jsz-hperp}, which can only be explained when  the contribution of states outside the magnetic gap is also considered. Nevertheless, such analogy does give an intuitive argument to qualitatively justify the origin of the equilibrium spin current. We shall thus briefly recall it here below.

\subsubsection{The analogy with QSH edge states and the role of the states inside the gap} 

To illustrate the analogy with the QSH helical states, let us consider a weak applied magnetic field, $\Delta_Z \ll 2 E_{SO}$, and compare Fig.\ref{Fig2}(b) with the spectrum in the absence of magnetic field depicted in Fig.\ref{Fig2}(a). One observes that,  while any energy $|E|\ge\Delta_Z$ outside the magnetic gap   always identifies two pairs of states  like in  Fig.\ref{Fig2}(a), for energies  $|E|<\Delta_Z$   inside the gap  only one pair  of ``time-reversed states"  is left [filled circles  of Fig.\ref{Fig2}(a) and \ref{Fig2}(b)]. It  thus seems at first  plausible that the spin current  is mainly due to the only pair remaining inside the gap, whose contribution at any energy $|E|<\Delta_Z$ is left unbalanced.
Notably,  such  pair surviving inside the gap consists of counterpropagating   states with equal and opposite velocities, so that it carries no net current. Furthermore, as the magnetic field is weak, their spin directions of  are essentially determined by the Rashba coupling, and are (almost) opposite to each other, similarly to the helical states of a Quantum Spin Hall (QSH) system. Explicitly, for $\alpha>0$  right-moving states have (almost) spin-$\uparrow$, while left-moving states have (almost) spin-$\downarrow$, whereas the  opposite occurs for $\alpha<0$. This analogy explains why the states inside the gap carry an equilibrium spin current, and why its flow is determined by the sign of $\alpha$, and not by the sign of $h_x$.\\

Although intuitive and appealing, the above analogy with the QSH edge states is not sufficient to account for the behavior of  the nanowire equilibrium spin current shown in Fig.\ref{Fig3-Jsz-hperp}. This
is straightforwardly seen  by focusing on the case of chemical potential lying in the middle of the magnetic gap ($\mu=0$), described by thick solid black curve of Fig.\ref{Fig3-Jsz-hperp}. It reveals two key features. 
The first one is the sign of the spin current:  the   analogy with QSH would predict that at weak fields the spin current has the same sign $\mbox{sgn}(\alpha)$ as the RSOC, whereas the actual sign is quite the opposite, as described by the following asymptotic expression at weak fields
{\small
\begin{eqnarray}
\left.  J^s_z \right|_{\mu=0} 
    \simeq   - \displaystyle\frac{\mbox{sgn}(\alpha) }{8\pi} \frac{\Delta_Z^2}{E_{SO} }  \left( \ln \left(\frac{8 E_{SO} }{{\Delta_Z}}\right)- 1   \right)  \hspace{0.4cm}   \Delta_Z \ll E_{SO}\,\, . \hspace{0.4cm}
\end{eqnarray} 
}

\noindent and illustrated in the case $\alpha>0$ by the {\it negative} thick solid black curve.
The second feature   is the behavior at strong magnetic field: When the Zeeman-dominated regime $\Delta_Z > 2 E_{SO}$  is entered [see Fig.\ref{Fig2}(c)], the  magnetic field tends to align parallely along~$x$ the spins of the two states at the Fermi level, and the analogy with the QSH helical pair is lost. One would thus expect  the spin current to decrease with~$\Delta_Z$, whereas the solid black curve of Fig.\ref{Fig3-Jsz-hperp} shows that the magnitude of $J^s_z$ is {\it increasing}, as confirmed by the strong field asymptotic behavior
{\small
\begin{eqnarray}\label{asym-2}
\left.  J^s_z \right|_{\mu=0}  \simeq \displaystyle - \frac{\mbox{sgn}(\alpha)}{\pi}    \left(\frac{\sqrt{ \Delta_Z E_{SO}}}{3}  -   \frac{E_{SO}^{3/2}}{5\sqrt{\Delta_Z}}\right) \hspace{0.3cm}   \Delta_Z\gg E_{SO}.\hspace{0.5cm} 
\end{eqnarray} 
}
\noindent 
This proves that  the equilibrium spin current arising in the nanowire persists far beyond the Rashba-dominated regime, and that the analogy with QSH edge states is not sufficient to account for the behavior displayed in  Fig.\ref{Fig3-Jsz-hperp}, not even qualitatively.

\subsubsection{The  role of the states outside the gap} 
In order to go beyond the argument of the  analogy with the QSH edge states and  explain the behavior of the equilibrium spin current   in  Fig.\ref{Fig3-Jsz-hperp}, two  aspects must be taken into account. First, while the charge current  is essentially due to the states near the Fermi level, the {\it equilibrium} spin current takes contribution from {\it all} states up to the Fermi level~$\mu$. 
Second, the states with energy $|E|\ge\Delta_Z$ outside the gap turn out to provide a significant contribution to the spin current. 
This seems counter-intuitive at first, since a comparison of the electron spectra without magnetic field [Fig.\ref{Fig2}(a)] and with magnetic field [Fig.\ref{Fig2}(b)] would   suggest that the difference causing the spin current  stems from the states inside the gap only. However, when both RSOC and magnetic field are present in the nanowire,  spin and velocity are not factorizable, $\langle \hat{S}_z v \rangle \neq \langle \hat{S}_z\rangle \langle v\rangle$, implying that the spin current contribution of a state cannot be simply deduced  by separately inspecting its spin direction  (spin arrows in the spectrum) and its group velocity   (slope of the spectrum). This is why, even in the Rashba-dominated regime, the scenario  for the nanowire is more complicated  and richer than in QSH edge state case. 
 More specifically, for each  eigenstate $\Psi_{k\pm}(x)=w_{k\pm}e^{i k x}$ with wavevector $k$ in the band $b=\pm$,  the following inequality holds
\begin{eqnarray}\label{inequality}
 \langle \hat{S}_z  v \, \rangle_{k\pm} -  \langle \hat{S}_z  \rangle_{k\pm} \, \langle v \rangle_{k\pm}   =-\frac{\alpha}{2} \frac{h_x^2}{(\alpha k)^2+h_x^2}   \neq 0
\end{eqnarray}
where $\langle \ldots \rangle_{k\pm} =\langle \Psi_{k\pm} |\ldots | \Psi_{k\pm} \rangle$.  In particular, for states with a large wavevector $|k| \sim  k_{SO}$, the right-hand side of Eq.(\ref{inequality}) is small in the limit $h_x \ll E_{SO}$, so that spin and velocity effectively factorize, similarly to the QSH edge states.    
In contrast, for states with small wavevector $|k| \ll k_{SO}$, the lack of factorization becomes important, and spin current gets a non-trivial contribution.
This is particularly striking for the two states at $k=0$: While their group velocity vanishes for any arbitrarily small magnetic field  as they become the extremal points of the two bands at the gap edges [$\langle v  \rangle_{0,\pm}=0\,\, \forall {h_x} \neq 0$, see Fig.\ref{Fig2}(b)], their spin current contribution  is insensitive to $h_x$ and remains equal to $-\alpha/4\pi$, i.e. the value in the absence of magnetic field  [see Eq.(\ref{Jsz-hperp})]. At the same time, because the density of states (DOS) $g_\pm(E)$ of the 1D nanowire  diverges  at the gap edges $E=\pm \Delta_Z$ where the group velocity vanishes,  the contribution of the  $k=0$ states to  the total spin current turns out to be dramatically enhanced by   the presence of the magnetic field. Note that this is quite different from the behavior of the charge current, where the DOS divergence is compensated by the vanishing of the velocity $v$. 

This consequence of the inequality (\ref{inequality})  --a spin current carried by states with vanishing velocity--  deserves a comment. It is worth noting that  Eq.(\ref{inequality}) originates from the fact that, when both RSOC and magnetic field are present, the nanowire eigenstates are {\it not} eigenstates of the velocity operator (\ref{vel-op-def}). Rather, they are linear combinations $|{k\pm}\rangle =w_{k\pm, \uparrow} |k\!\uparrow\rangle + w_{k\pm, \downarrow} |k\!\downarrow\rangle$ of states with opposite spins and propagating in opposite directions  [see dashed curves of Fig.\ref{Fig2}(b)]. In particular,  at $k=0^+$ and $k=0^-$, the linear combinations involve states with exactly opposite velocities, opposite spins, and the same spin current. This is why the $k=0$ states carry a finite  spin current   despite a vanishing net velocity.
For this reason, already at the band edges $E=\pm \Delta_Z$ the dominating contribution to the spin current counterintuitively arises  from the states at $k=0$, and not from the other pair of states at the same energy and with finite wavevector. 
More in general, each energy  $|E|\ge \Delta_Z$ outside the gap identifies {\it two} pairs of counterpropagating states,  one characterized by a large and one by a small wavevector,  carrying spin currents of opposite signs. While for vanishing magnetic field a perfect cancellation occurs, in the presence of the magnetic field the spin current contribution of the pair with small wavevector prevails.

\subsubsection{Spectral decomposition of the equilibrium spin current} 
It is useful to introduce its spectral decomposition of the equilibrium spin current Eq.(\ref{Jsz-hperp}) through 
\begin{equation} \label{spectral-decomposition}
J^s_z =   \int dE \left( \sum_{b=\pm}J^s_{z,b}(E)\,g^{}_b(E) \right)f^\circ(E)\quad,
\end{equation}
where $g_\pm(E)=(2\pi)^{-1}   \int   dk \, \delta(E-E_\pm(k))$ denotes the DOS,
and 
\begin{eqnarray}
J^s_{z,\pm}(E)\,g^{}_\pm(E) &=& \displaystyle  \int   \frac{dk}{2\pi}\, \delta(E-E_\pm(k)) \, J^s_{z,\pm}(k)\,\,, \label{spectral-weight}
\end{eqnarray}
are the spectral weights related to the two bands $b=\pm$, illustrated by the two thin solid curves in Fig.\ref{Fig2}(b) and (c),  in the case $\alpha>0$ [the reader should refer to the left vertical axis (energy) and to the upper horizontal axis (spectral weight of spin current)]. 
Let us focus in particular on the {\it sign} of the spectral weights of the two bands. As one can see, independently  of whether the nanowire is in the Rashba-dominated or in the Zeeman-dominated regime [see Fig.\ref{Fig2}(b) and (c)], all states of  the upper band $E_{+}$ carry a spectral weigth $J^s_{z,+}(E)\,g^{}_+(E)$ with a sign $-\mbox{sgn}(\alpha)$ (i.e. negative in the case $\alpha>0$ shown in Fig.\ref{Fig2}), with a van Hove divergence  appearing in correspondence of the upper band edge  $E= \Delta_Z$. In contrast, the lower band $E_{-}$ exhibits states both with positive and negative spectral weight, whose energy behavior depends on the specific regime. In the Rashba-dominated regime, the spectral weight  $J^s_{z,-}(E)\,g^{}_-(E)$ consists of two branches: the states with energy $E>-\Delta_Z$ (including the ``helical states" at $E=0$) carry a spectral weight that has a sign $+\mbox{sgn}(\alpha)$ and is weakly dependent on energy, whereas the states below the band edge $-E^\prime_{min} \le E \le -\Delta_Z$ carry a spectral weight that has a sign $-\mbox{sgn}(\alpha)$ and is strongly energy dependent. In particular, a van Hove singularity arises as $E \rightarrow -\Delta_Z$, due to the states with small wavevector $|k|\ll k_{SO}^\prime$. Notice that a third van Hove singularity is present at the local minima (\ref{Eminprime-Rashba}) in Fig.\ref{Fig2}(b). This unbalance in the sign of the spectral weight shows that low energy states play a major role in  determining the sign and magnitude of the equilibrium spin current, which cannot be deduced just from the states near $E \simeq 0$.

\subsection{Dependence of the equilibrium spin current on the magnetic field and chemical potential}
With the help of the spectral decomposition introduced above, we can now understand the behavior of the spin current shown in Fig.\ref{Fig3-Jsz-hperp}. Indeed, as can be seen from  Eq.(\ref{spectral-decomposition}), the zero temperature spin current  is obtained by integrating the spectral weight up to  the chemical potential $\mu$,  which thus determines which branches --and therefore which sign-- of the spectral weight are active.

In particular, at $\mu=0$ only the lower band $E_{-}$ is involved (see Fig.\ref{Fig2}). The branch of the spectral weight $J_{z,-}^s(E) g_{-}(E)$ that has sign $-\mbox{sgn}(\alpha)$ (i.e. negative  in Fig.\ref{Fig2}) and contains the van Hove singularities  dominates over the 
branch with sign $+\mbox{sgn}(\alpha)$. This explains why the spin current at $\mu=0$, described in the case $\alpha>0$ by the solid black curve of Fig.\ref{Fig3-Jsz-hperp}, is negative, oppositely to what one would naively conclude from the analogy with QSH edge states.
The inequality (\ref{inequality}) also justifies why the spin current  increases in magnitude with the magnetic field.  Indeed, despite a strong field $h_x>2E_{SO}$  [see Fig.\ref{Fig2}(c)] tends to align spins along the direction $x$, so that $\langle \hat{S}_z \rangle_{k\pm} \rightarrow 0$, the first term on the l.h.s. of (\ref{inequality}) remains finite. In fact,  the range of states $\Psi_{k-}$ of the lower band contributing to the spin current $ \langle   \hat{S}_z \, v\rangle_{k\pm}$ with a sign $-\mbox{sgn}(\alpha)$ increases with $h_x$. Explicitly, such range is identified by $|k|<k^*$, where
\begin{equation}
\label{kstar} k^*=k_{SO} \sqrt{\frac{1+\sqrt{1+h_x^2/E_{SO}^2}}{2}}\quad,
\end{equation}
or equivalently by the energy window $E^\prime_{min}<E<-E^*$, with $E^*\doteq -E_{-}(k^*) \equiv \hbar^2 {k^*}^2/2m^*$. In particular, in the Zeeman-dominated regime [see Fig.\ref{Fig2}(c)], where $E^\prime_{min}$ is given by Eq.(\ref{Eminprime-Zeeman}), such range increases with $h_x$, and so does the magnitude of the spin current.

In the case $\mu<0$, the spin current is described by dashed red curve of Fig.\ref{Fig3-Jsz-hperp}. Although only the lower band $E_{-}$ contributes to $J^s_z$ at zero temperature, two situations can be identified in this case. For weak fields $\Delta_Z< |\mu|$, only   the branch with sign $-\mbox{sgn}(\alpha)$ of the spectral weight $J_{z,-}^s(E) g_{-}(E)$ is filled  [see thin solid curves of Fig.\ref{Fig2}(b)]. At $\Delta_Z=|\mu|$, the van Hove singularity of the spectral weight at $E=-\Delta_Z$ is encountered, whereas for higher field values the branch $J_{z,-}^s(E) g_{-}(E)$ with sign $+\mbox{sgn}(\alpha)$ starts to contribute. This explains why   the dashed red curve of  Fig.\ref{Fig3-Jsz-hperp} displays a cusp at $\Delta_Z=|\mu|$ and increases in magnitude with a lower rate for $\Delta_Z > |\mu|$. At high fields, the behavior is qualitatively similar to the case of $\mu=0$.  

For $\mu>0$ the spin current is given by the thin blue solid and green dash-dotted curves in Fig.\ref{Fig3-Jsz-hperp}. For weak fields $\Delta_Z< \mu$ both bands $E_{-}$ and $E_{+}$ contribute to $J^s_z$. However, as $h_x$ is ramped up, part of the spectral weight $J_{z,+}^s(E) g_{+}(E)$, which carries a sign $-\mbox{sgn}(\alpha)$,  is cut off. The  spin current for $\mu>0$ at low fields thus increases with $\Delta_Z$. However, when $\Delta_Z=\mu$, the van Hove singularity of the upper band edge is encountered, the upper band is pushed above the Fermi level and does not contribute anymore. This is why a cusp occurs at $\Delta_Z= \mu$  in Fig.\ref{Fig3-Jsz-hperp}. For higher fields   the states of the spectral weight $J_{z,-}^s(E) g_{-}(E)$ carrying a sign $-\mbox{sgn}(\alpha)$ dominate and the spin current becomes negative.\\

\begin{figure} 
\centering
\includegraphics[width=\columnwidth]{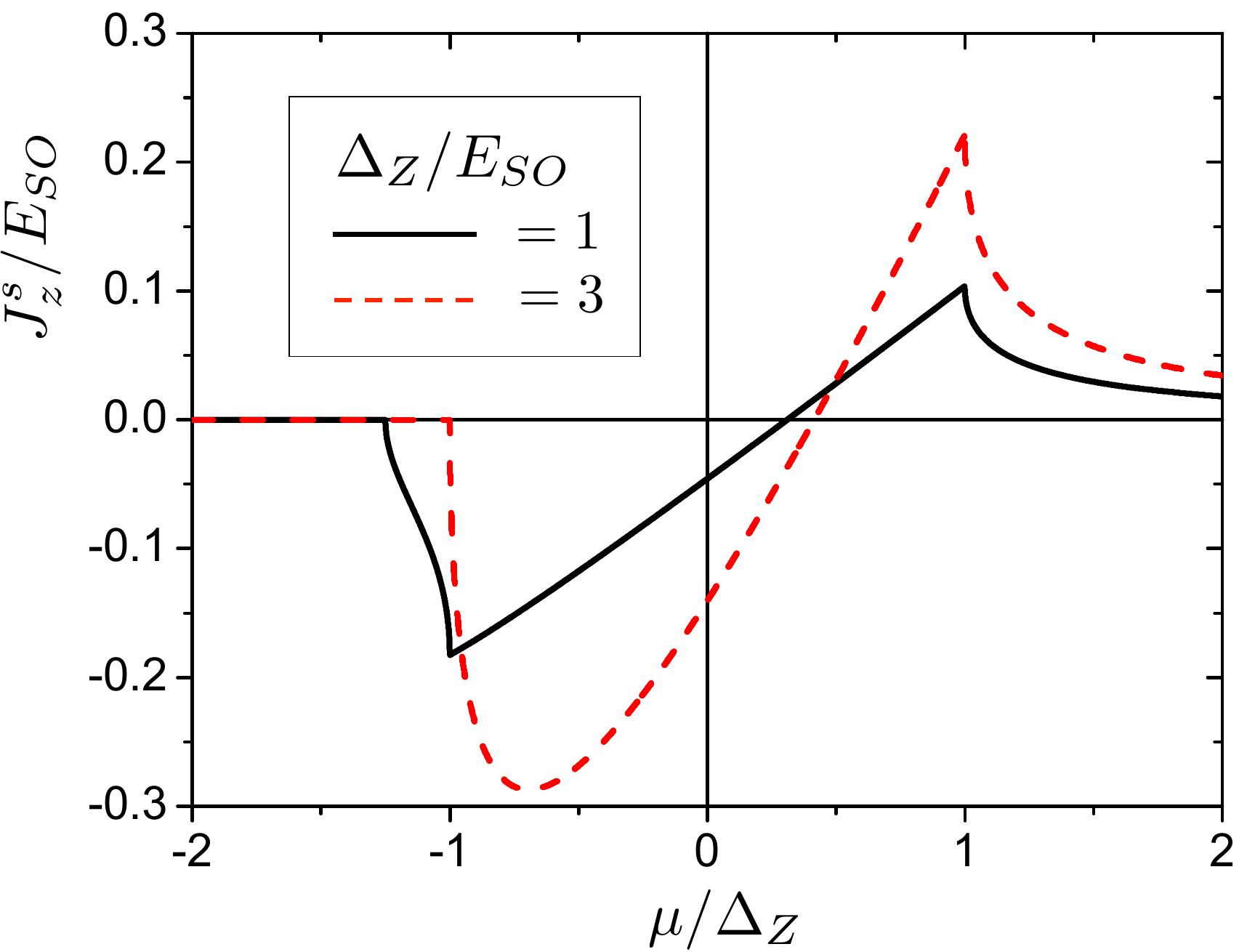}
\caption{(Color online) The zero temperature behavior of the equilibrium spin current $J^s_z$ as a function of the chemical potential $\mu$ of the nanowire, for two values of the Zeeman gap energy $\Delta_Z$ related to the magnetic field applied along the nanowire axis. The solid black curve refers to the Rashba-dominated regime $\Delta_Z/E_{SO}=1$ [see Fig.\ref{Fig2}(b)], while the dashed red curve to the Zeeman-dominated regime $\Delta_Z/E_{SO}=3$ [see Fig.\ref{Fig2}(c)]. Singularities correspond to the gap edges $\mu=\pm \Delta_Z$}
\label {Fig4-Jsz-mu}
\end{figure}

Importantly, these results also show that the equilibrium spin current can   be tuned by the chemical potential $\mu$ of the nanowire, which can be  varied e.g. with an applied gate. This is
 shown in Fig.\ref {Fig4-Jsz-mu}, which displays $J_z^s$ as a function of $\mu$, for two different values of magnetic field. The solid curve refers to the Rashba-dominated regime $\Delta_Z< 2 E_{SO}$  [see Fig.\ref{Fig2}(b)], and the spin current increases (in magnitude) as $\mu$ varies from the minimal energy of the band $E_{min}^\prime$ [see Eq.(\ref{Eminprime-Rashba})] to the value $\mu=-\Delta_Z$ corresponding to the first gap edge, where $J^s_z$ exhibits a cusps and starts to decrease in magnitude. Then, inside the magnetic gap, $|\mu|<\Delta_Z$, $J^s_z$ grows essentially linearly with $\mu$. Interestingly, the change of sign occurring with varying $\mu$ shows that the polarization of the spin current can be controlled through the chemical potential.
 A~second cusp then arises at the other gap edge $\mu=+\Delta_Z$ where the states of the upper band $E_{+}$ start to provide an opposite contribution to the spin current. Eventually,  for $\mu \gg \Delta_Z$, $J^s_z$ decreases since at such high energies the effect of the magnetic field becomes negligible. In contrast, in the case of the Zeeman-dominated regime $\Delta_Z> 2 E_{SO}$, illustrated by the dashed curve of Fig.\ref {Fig4-Jsz-mu}, the first cusp reverses its curvature and occurs at the lower edge $\mu=-\Delta_Z$, corresponding to the minimum of the lower band at $k=0$ [see Fig.\ref{Fig2}(c)].   \\

\section{Inhomogeneous Rashba profile: nanowire contacted to the leads}
\label{sec-4}
As observed in the Introduction, in realistic setups the nanowire has a finite length $L_w$ and is typically contacted to metallic electrodes, where the Rashba spin-orbit coupling is negligible, as sketched in Fig.\ref{Fig-setup}(b).  We shall thus now investigate the equilibrium  spin properties for the system nanowire+leads.
Inspired by the model used in Ref.[\onlinecite{sanchez_2008}] to analyze transport properties, we shall adopt in Eq.(\ref{H-realspace}) an inhomogeneous Rashba coupling, which takes a finite value  in a central region (the nanowire), and vanishing value in the external regions (the leads). However, instead of the abrupt piecewise constant profile used in Ref.[\onlinecite{sanchez_2008}], here we take into account a smooth crossover between the leads and the nanowire.
Explicitly, we shall adopt
{\small
\begin{equation}\label{alpha-profile}
\alpha(x)=\frac{\alpha_0}{2} \left[ {\rm Erf}\left(\frac{x+L_w/2}{\sqrt{2}\, \lambda} \right)+{\rm Erf}\left(\frac{L_w/2-x}{\sqrt{2}\, \lambda}\right)\right]
\end{equation}
}

\noindent where $\alpha_0$ denotes the value of the Rashba coupling in the ``bulk" of the nanowire, $L_w$ the nanowire length and  $\lambda$   the crossover lengthscale. The magnetic field is assumed to be uniform over the whole system (nanowire + leads), for which we can adopt periodic boundary conditions over a total   length $L$ of the system nanowire+leads, without loss of generality.

A crucial difference with respect to the homogeneous case analyzed in Sec.\ref{sec-3} is that the inhomogeneous profile~(\ref{alpha-profile}) makes the Hamiltonian (\ref{H-realspace}) not commute with~$p_x$. Nevertheless, the states $\Psi_{k\uparrow,\downarrow}(x)=e^{i k x} \chi_{\uparrow,\downarrow}$, obtained by the product of plane waves and  the eigenvectors $\chi_\uparrow=(1,0)^T$ and $\chi_\downarrow=(0,1)^T$ of the Rashba spin direction $\sigma_z$, form a complete set for the Hilbert space. Exploiting Eq.(\ref{Psi-Fourier}), the  Hamiltonian (\ref{H}) in such basis  reads 
\begin{eqnarray}\label{H-k-inhomo}
\hat{\mathcal{H}}  &= &\sum_{k_1,k_2} \hat{\mathcal{C}}^\dagger_{k_1} \left( \left( \varepsilon^0_{k_1}\sigma_0- \mathbf{h} \cdot \boldsymbol{\sigma}\right) \delta_{k_1,k_2} \, - \right. \nonumber \\
& & \left. \hspace{1cm} -\alpha_{k_1-k_2}\frac{k_1+k_2}{2}\, \sigma_3\right) \hat{\mathcal{C}}^{}_{k_2} \quad,
\end{eqnarray}
where
$
\alpha_q=L^{-1}\int  \alpha(x)\,e^{-i q x} dx$ 
denotes the Fourier component of the inhomogeneous RSOC. The homogeneous case is recovered by $\alpha_{k_1-k_2}=\alpha\, \delta_{k_1,k_2}$, while for the profile (\ref{alpha-profile}) one has 
\begin{equation}
\alpha_q= \frac{2 \alpha_0}{qL_w} \sin\left(\frac{q L_w}{2} \right)\, e^{-(q \lambda)^2/2}\quad.
\end{equation} 
We have thus numerically diagonalized the Hamiltonian matrix Eq.(\ref{H-k-inhomo}), thereby obtaining diagonalizing operators $\hat{c}_\xi$'s. Then, by expressing the operators (\ref{charge-density-def}), (\ref{charge-current-def}), (\ref{spin-density-def}), (\ref{spin-current-def}), (\ref{Th-def}) and (\ref{TSO-def})  in terms of the $\hat{c}_\xi$'s, the equilibrium  expectation values (\ref{equil-exp-val}) are evaluated by using $\langle \hat{c}^\dagger_\xi \hat{c}^{}_\xi \rangle_\circ=f^\circ(E_\xi)$. 
Differently from the bulk limit of Sec.\ref{sec-3}, in this case the inhomogeneity of the RSOC makes the equilibrium expectation values space-dependent, as we shall discuss here below.

\subsection{Orthogonal spin polarization and spin torque pinned at the interfaces}
\label{sec-4A}
\begin{figure} 
\centering
\includegraphics[width=\columnwidth]{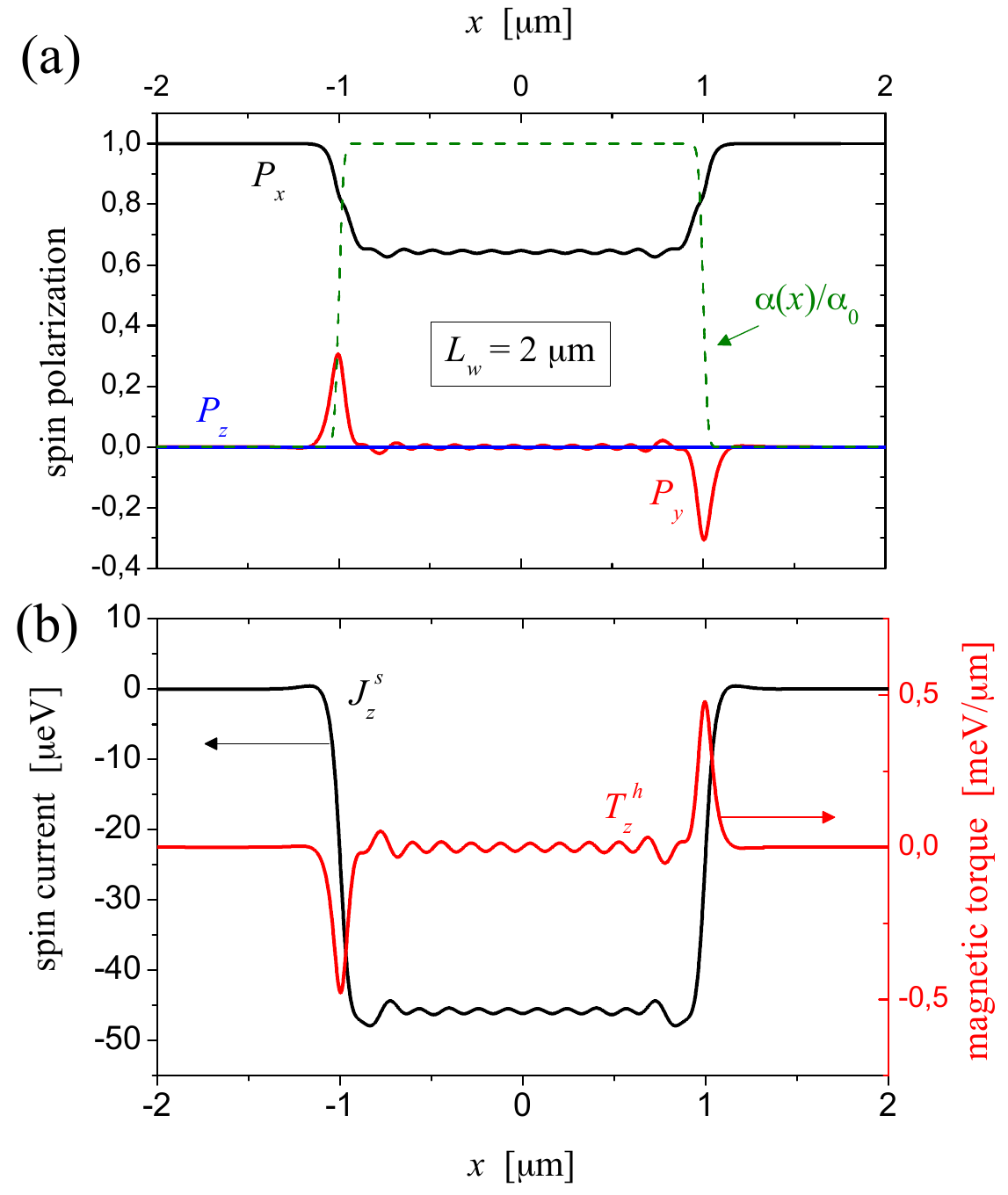}
\caption{\label{Fig-inhomo-long} (Color online) Equilibrium spin properties of a long nanowire ($L_w=2 \mu{\rm m}$) contacted to leads, with $\Delta_Z=0.4\,{\rm meV}$, $\mu=-0.2 \,{\rm meV}$, at  the temperature $T=50 \,{\rm mK}$. In the inhomogeneous RSOC profile Eq. (\ref{alpha-profile}), the smoothing length is $\lambda=20\,{\rm nm}$, while the value of $\alpha_0$ corresponds to a Rashba energy $E_{SO}=0.3 \,{\rm meV}$ in the nanowire bulk, through Eq.(\ref{ESO-def})  with an effective mass $m^*=0.015 m_e$. \\ (a) The solid curves describe the space profile of the three components of the spin polarization (\ref{P-def}), while the dashed curve depicts the inhomogeneous RSOC profile. At the interfaces with the leads, a  spin polarization $P_y$ orthogonal to the substrate plane appears.  (b) The solid black curve describes the space profile of the spin current (to be read on the left vertical axis), while the solid red curve describes the spin torque  (to be read on the right vertical axis).}
\end{figure}
\begin{figure} 
\centering
\includegraphics[width=\columnwidth]{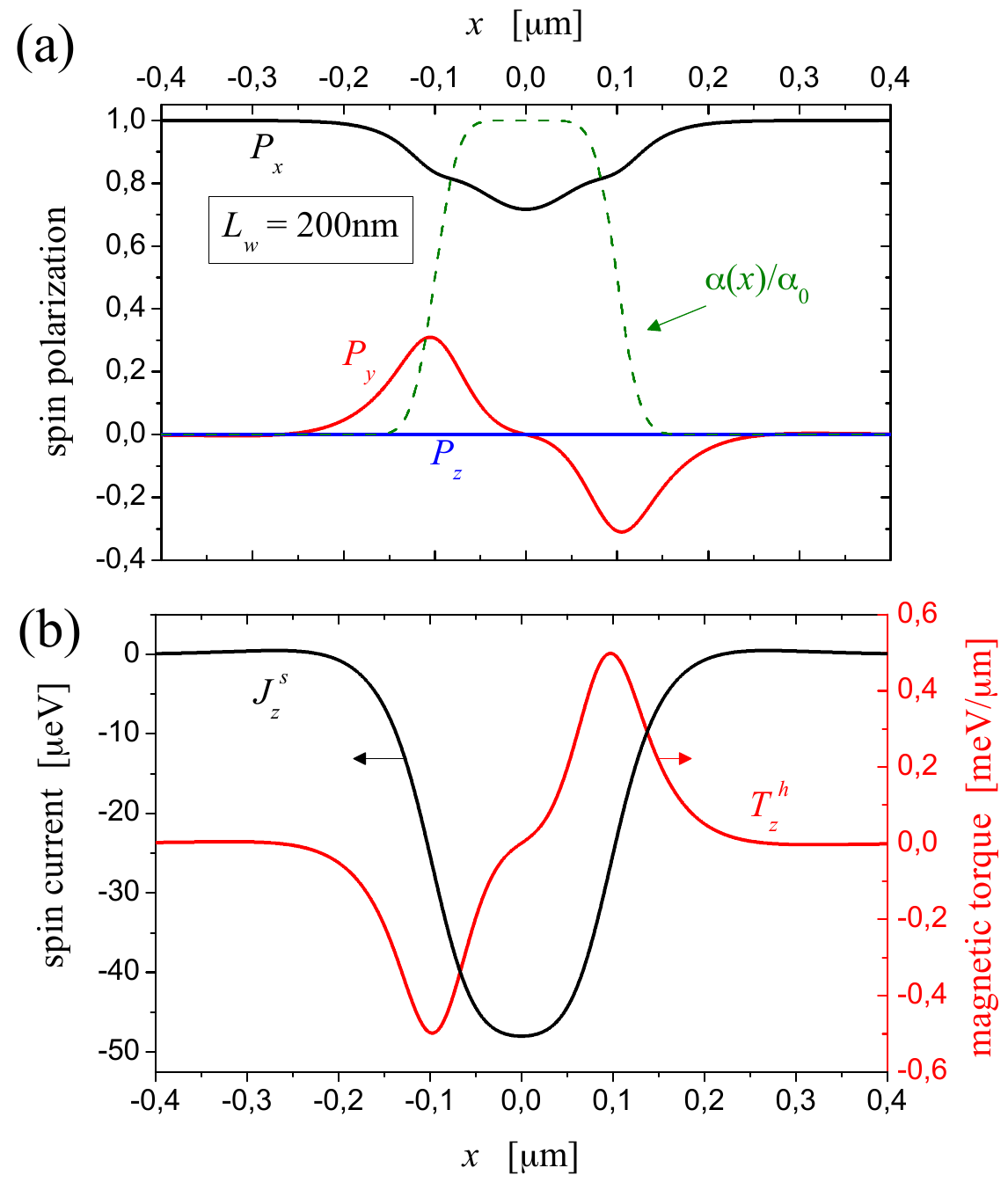}
\caption{\label{Fig-inhomo-short} (Color online) Equilibrium spin properties of a short nanowire ($L_w=200{\rm nm}$) contacted to leads. All other parameter values and the meaning of the curves are the same as in Fig.\ref{Fig-inhomo-long}. }
\end{figure}

The results  for the spin sector of the nanowire contacted to leads are shown in Figs.~\ref{Fig-inhomo-long} and~\ref{Fig-inhomo-short},  and refer to  the cases of a long ($L_w=2 \mu{\rm m}$), and a short ($L_w=200 {\rm nm}$) nanowire, respectively. The parameters correspond to experimentally realistic values for Rashba nanowires, with a smoothing length $\lambda=20{\rm nm}$ of the Rashba profile (\ref{alpha-profile}).  

Let us focus first  on the long nanowire. Figure~\ref{Fig-inhomo-long}(a) displays the profile of inhomogeneous RSOC (dashed green curve) and the  behavior of the three components of the spin polarization  $\mathbf{P}$ as a function of the space coordinate $x$ along the nanowire, while Fig.\ref{Fig-inhomo-long}(b) shows the spin current $J^s_z$ and the spin torque $T^h_z$ along the Rashba direction $z$ (the other spin current components $J^s_{x,y}$ and the spin-orbit torque ${\bf T}^{SO}$ vanish). In the bulk of the nanowire ($|x|<1\mu{\rm m}$), one  observes a weakly oscillatory behavior of the 
spin polarization component~$P_x$, collinear with the  applied magnetic field ${h_x}$   [black curve of Fig.\ref{Fig-inhomo-long}(a)] and of the spin current component $J^s_z$ polarized along the Rashba field direction $z$ [black curve of Fig.\ref{Fig-inhomo-long}(b)]. These oscillations, which describe deviations from   the   bulk values of $P_x$ and $J_z^s$ obtained from the analysis of the infinitely long wire (in Sec.\ref{sec-3}),  are a quantum interference effect due to the finite size of the nanowire, which leads the states of the  nanowire bulk to be reflected at the interfaces, for each energy $E$. The wavelength of these interferences can   be estimated as
\begin{equation}\label{lambda-osc-def}
\lambda_{osc}\simeq \frac{\pi \hbar}{\sqrt{2 m^*}} \left(\mu+2 E_{SO}+\sqrt{4 \mu E_{SO}+4 E_{SO}^2+\Delta_Z^2}\right)^{-1/2}
\end{equation}
and depends on the Rashba spin-orbit energy, chemical potential and magnetic gap energy. For the specific values of Fig.\ref{Fig-inhomo-long} it takes a value of about $165{\rm nm}$.

Furthermore, the space dependence of the polarization $P_x$ collinear to the magnetic field [black curve  of Fig.\ref{Fig-inhomo-long}(a)] displays a crossover from the ``nanowire value" (due to the magnetic field but also affected by the Rashba coupling) to a ``lead value" (bigger in magnitude and purely due to the magnetic field).  Importantly, at the nanowire/lead interfaces, one observes the appearance of   a spin polarization $P_y$ {\it orthogonal} to both the direction $x$ of the applied magnetic field $\mathbf{h}=(0,0,h_x)$ and   the Rashba direction $z$.
Note that  $P_y$ takes opposite signs at the two interfaces, as shown by the red curve of Fig.\ref{Fig-inhomo-long}(a).  
This behavior can be understood as follows: at each energy $E$ the wavefunction  inside the nanowire is a superposition of states whose spin is lying in the $x$-$z$ plane, as a result of the combined effect of the magnetic field (pointing along $x$)  and the  effective Rashba field (pointing along $z$). 
In contrast,  in the bulk of the leads, where no RSOC is present, the spin  is simply oriented like the magnetic field  along~$x$.  At the  interfaces,  the wavefunction matching between these two regions can occur  only if an orthogonal spin component $\langle S_y\rangle$ arises. This is causes the behavior of the  spin polarization $P_y$ in Fig.\ref{Fig-inhomo-long}(a).

In turn, precisely because such interface polarization is orthogonal to the magnetic field $\mathbf{h}$, it generates a spin torque $\mathbf{T}^h= \rho (\mathbf{P}  \times \mathbf{h})=(0,0,T^h_z)$, as can be straightforwardly deduced from from Eqs.(\ref{spin-density-def}), (\ref{Th-def}) and (\ref{P-def}). Thus, the spin torque component 
\begin{equation}\label{Tz-to-Sy}
T^h_z=- P_y\, \rho \, h_x     \quad,
\end{equation}
is  also pinned at the interfaces, as shown in the red curve of Fig.\ref{Fig-inhomo-long}(b)].
 
At the same time, across the interface we also observe a suppression of the equilibrium spin current $J^s_z$ [black curves  of Fig.\ref{Fig-inhomo-long}(b)], which  eventually vanishes  in the bulk of the leads, consistently with the result shown in Sec.\ref{sec-3} that $\mathbf{J}^s$ vanishes when the RSOC is absent. The space profile of the $J^s_z$ is closely related to the interface spin torque $T^h_z$, according to the equation
\begin{equation} \label{cont-eq-2}
\partial_x  \mathbf{J}^s=\, \mathbf{T}^{h} \quad,
\end{equation}
showing that any space variation of the spin-current $\mathbf{J}^{s}=(0,0,J^s_z)$  is associated with the presence of a spin torque $\mathbf{T}^{h}=(0,0,T^{h}_z)$. 
This can be seen by taking the expectation values  (\ref{equil-exp-val}) of the  equation of motion~(\ref{spin-cont-eq}), and by noting that at equilibrium $\mathbf{S}$ is time-independent and $\mathbf{T}^{SO}=0$ for symmetry reasons.\\

For a shorter nanowire, $L_w=200 {\rm nm}$, the behavior of these quantities  is shown in Fig.~\ref{Fig-inhomo-short}. Notice that in this case,  where the Rashba profile (green dashed curve) is relatively smooth,  the  lengthscale (\ref{lambda-osc-def}) of the oscillations becomes comparable with the length $L_w$ of the nanowire, so that $P_x$ and $J^s_z$ [black curves in Fig.\ref{Fig-inhomo-short}(a) and (b), respectively] do not exhibit the oscillations  observable in the long nanowire case of Fig.\ref{Fig-inhomo-long}, and the only residual track of the bulk value of $P_x$ and $J^s_z$ is the minimum  located at the center of the nanowire.
Furthermore,   the orthogonal spin polarization component $P_y$ and the spin torque $T^h_z$ [red curves in Figs.\ref{Fig-inhomo-short}(a) and \ref{Fig-inhomo-short}(b), respectively] exhibit a node in the middle of the nanowire and one sign  change across the two interfaces.\\

\subsection{Equilibrium spin current from spin bound states}
\label{sec-4-B}
The results obtained for the inhomogeneous RSOC also provide an alternative way to interpret the equilibrium spin current of the nanowire from purely a spin polarization argument. To illustrate it, let us  focus for simplicity on the helical-like states inside the magnetic gap. In  the nanowire  the spin direction is mainly determined by the RSOC and is pointing along $z$, whereas in the leads it is dictated only by the magnetic field applied along $x$. Thus, in a  scattering formalism picture, an electron with (almost) spin-$\uparrow$ traveling rightwards from the nanowire bulk cannot freely propagate into the right lead, which instead acts as a magnetic ``barrier",  backscattering the electron at the interface  into a left-moving electron  with (almost) spin-$\downarrow$. The appearance of the interface spin torque $T^h_z$ [red curves of Figs.\ref{Fig-inhomo-long}(b) and \ref{Fig-inhomo-short}(b)] is the hallmark of such spin-flip process.  
A similar process occurs at the left interface, where left-moving electrons with (almost) spin-$\downarrow$ are backscattered into right-moving electrons with (almost) spin-$\uparrow$. Sandwiched between the two magnetic ``barriers", i.e. the leads, the nanowire hosts {\it spin bound states}, characterized by  electrons with opposite spins traveling  in opposite directions, which are converted into each other by the spin-torques present at the interfaces. These bounds states, sketched in Fig.\ref{Fig-setup}(b), carry the equilibrium spin current inside the wire. Notice that the presence of the magnetic field inside the nanowire is crucial in creating the gap and removing the other   pair of counterpropagating states with opposite spin orientations: that would otherwise give rise to a spin torque opposite to the previous one, at each interface, and to another bound state inside the nanowire cancelling the above spin current.
For energies outside the gap, the reasoning follows the same lines, the bound states being formed out of two inequivalent pairs of states, though.

These spin bound states are reminiscent of the Andreev bound states (ABS) of a superconductor-normal-superconductor hybrid junction, where electrons (holes) are back-reflected as holes (electrons) at each interface, carrying a net charge current.  
However, differently from charge current, which is conserved across the interface transforming from a ABS current in the normal region into a superfluid current in the superconducting leads, the spin current is not conserved and it vanishes in the leads, where carriers with opposite velocities have the same spin. The spin torques, which have opposite signs at the two interfaces, thus act  as a  ``source" and ``sink" of spin current,  as described by Eq.(\ref{cont-eq-2}).

\begin{widetext}
\begin{figure*}[b]
\centering
\includegraphics[width=18cm]{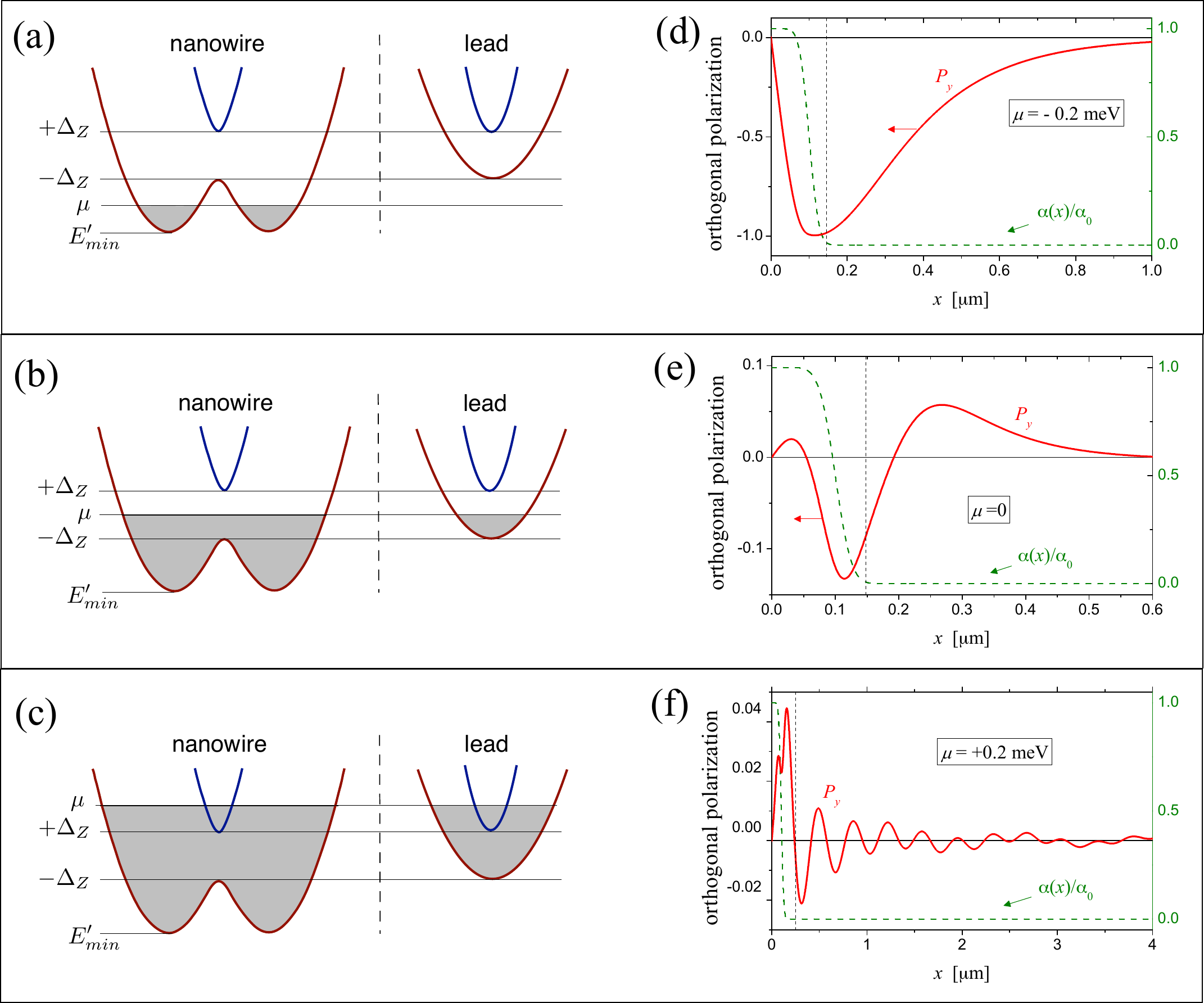}
\caption{\label{Fig7-interface} (Color online) The  penetration of the orthogonal spin polarization $P_y$ into a lead, caused by the inhomogeneous RSOC. Panels (a) (b) and (c)  sketch the ``band" spectrum of the bulk of the nanowire and the bulk of the lead, describing the three possible cases of the nanowire/lead interface in terms of the common chemical potential $\mu$.  Correspondingly, the solid curve in panels (d), (e) and (f) describe the space profile of $P_y$, while the dashed curve corresponds to the inhomogeneous Rashba profile (\ref{alpha-profile}) describing a short nanowire ($L_w=200{\rm nm}$) contacted to leads. The value of the Zeeman gap energy is $\Delta_Z=0.1\,{\rm meV}$, while in the inhomogeneous RSOC profile Eq.(\ref{alpha-profile}) the smoothing length is $\lambda=20\,{\rm nm}$, and the value taken for $\alpha_0$ corresponds to a Rashba energy $E_{SO}=0.3 \,{\rm meV}$ in the nanowire bulk. }
\end{figure*}
\end{widetext}

\subsection{Orthogonal spin penetration length inside the leads}
Figures \ref{Fig-inhomo-long} and \ref{Fig-inhomo-short} show that the orthogonal spin polarization  $P_y$ and the related spin torque $T^h_z$ in Eq.(\ref{Tz-to-Sy}) localized at the interface, penetrate inside the leads, even into regions where the RSOC is vanishing. 

The origin of this effect  is    the wavefunction matching at the nanowire/lead interfaces   mentioned in Sec.\ref{sec-4A}, which we now want to discuss  in more detail from the lead side. There,   only the magnetic field component $h_x$ is present, and at each energy $E$  the eigenfunction $\Psi$ of the leads
consists of a superposition of waves (propagating or evanescent), whose spinorial parts $w$ are eigenstates $w_\pm=(1,\pm1)^T/\sqrt{2}$  of~$\sigma_x$, related to the different eigenvalues $\pm 1$. For each of such eigenstates the expectation value of $\sigma_y$ is vanishing. However, in order for the wavefunction in the lead   to match the one in the nanowire, whose spin also has a component along $z$, {\it both} eigenstates $w_\pm$, must be present in $\Psi$ in the leads. It is their interference term $\langle w_{+}\sigma_y w_{-}\rangle$ that yields a finite expectation value of the orthogonal spin polarization $P_y= \langle \Psi^\dagger \sigma_y \Psi\rangle/\langle \Psi^\dagger \Psi\rangle$ in the leads. 
This effect is reminiscent of the anomalous expectation value induced by  a superconductor into a normal  lead, where both particle and hole states must be combined in the latter to match the wavefunction in the former. For this reason the penetration of $P_y$  in the leads represents a sort of ``spin proximity effect". There is, however, an important difference: Here, the expectation value of the orthogonal polarization $P_y$ is vanishing in {\it both} the nanowire bulk and in the lead bulk. The penetration is thus  purely due to the interface, i.e. to the inhomogeneity of the RSOC, and each interface side can be considered as "proximizing" the other.

To determine the penetration length into the leads, let us focus on an interface between a nanowire in the Rashba-dominated regime [see Fig.\ref{Fig2}(b)] and a lead. The latter is by definition  always in the Zeeman-dominated regime [see Fig.\ref{Fig2}(c)], since $\alpha=0$. For definiteness, we shall consider here below the case of a short nanowire $L_w=200{\rm nm}$, with spin-orbit energy $E_{SO}=0.3{\rm meV}$ in its bulk and a Zeeman gap energy $\Delta_Z=0.1{\rm meV}$. Then, depending on the value of the common equilibrium chemical potential~$\mu$, there can be three possible cases, illustrated on the left hand side of Fig.\ref{Fig7-interface}.

(a) $E_{min}^\prime < \mu<-\Delta_Z$. In this case the chemical potential lies above the band bottom Eq.(\ref{Eminprime-Rashba}) of the nanowire, but below the band bottom of the lead [see  Fig.\ref{Fig7-interface}(a)], 
so that the eigenfunction in the lead at the energy $E$ is a superposition of two evanescent waves $w_{+} e^{- q_{E+} x}$ and $w_{-} e^{- q_{E-} x}$, where   $q_{E\pm}=\sqrt{2m^*(-E\mp \Delta_Z)}/\hbar$. The penetration length for $P_y$, originating from the interference between these two wave components, can be estimated by the minimal values for $q_{E\pm}$, and reads
\begin{equation}\label{lambdaprox-a}
\lambda_{pen}  \simeq \frac{\pi\hbar}{\sqrt{2m^*(-\mu +\Delta_Z)} -\sqrt{2m^*(-\mu -\Delta_Z)}}\quad.
\end{equation}
For the value $\mu=-0.2{\rm meV}$ the estimate of (\ref{lambdaprox-a}) yields $\lambda_{pen}\sim 680{\rm nm}$, and is in agreement with the behavior of $P_y$ shown by the solid red curve of Fig.\ref{Fig7-interface}(d). The green dashed curve  again describes the inhomogeneous RSOC profile near the interface. Notice that in this regime of chemical potential the electron density $\rho$ decays exponentially into the lead. Because $\rho$ appears in the denominator of the polarization (\ref{P-def}), the profile of the spin polarization $P_y$ exhibits an exponential enhancement of its penetration length as compared to the spin density $S_y$ appearing in the numerator. Indeed $S_y$ decays over a lengthscale of about $180\,{\rm nm}$~[\onlinecite{nota-penetration}].\\

(b) $|\mu|<\Delta_Z$. In this case the chemical potential lies inside the magnetic gap [see  Fig.\ref{Fig7-interface}(b)], and the lead eigenfunction  is a superposition of a propagating wave  $w_{+} e^{\pm i q_{E+} x}$ and an evanescent wave $w_{-} e^{- q_{E-} x}$. Their interference thus leads to a damped oscillatory behavior for $P_y$, as can be seen in  Fig.\ref{Fig7-interface}(e). In this case  the electron density in the lead does not decay, and the penetration depth for both the spin polarization $P_y$ and the spin density $S_y$ is given by
\begin{equation}\label{lambdaprox-b}
\lambda_{pen}  \simeq \frac{\pi\hbar}{\sqrt{2m^*(-\mu +\Delta_Z)}}
\end{equation}
For $\mu=0$ Eq.(\ref{lambdaprox-b}) yields a  value of  about $500{\rm nm}$, in agreement with the solid red curve of Fig.\ref{Fig7-interface}(e). Notice that, as compared to case (a), the lengthscale (\ref{lambdaprox-b}) is shorter than the penetration length (\ref{lambdaprox-a}) of the polarization $P_y$, but longer than the penetration length of $S_y$ in that case~[\onlinecite{nota-penetration}]. \\

(c) $\mu>\Delta_Z $. In this case the chemical potential lies above the magnetic gap, so that also in the lead there are only propagating waves $w_{+} e^{\pm i q_{E+} x}$ and $w_{-} e^{\pm i k_{E-} x}$. Their interference is thus a long range oscillatory behavior, as described by the solid red curve of Fig.\ref{Fig7-interface}(f). Notice that the $x$-axis has been widely extended purposely as compared to Figs.\ref{Fig7-interface}(d) and (e).

It should be mentioned that the orthogonal spin penetration is of course ultimately cut off by the spin diffusion length in the leads, which depends on the specific material and the spin relaxation rate of the lead~[\onlinecite{fert_2001}]. In the low temperature regime considered here and for narrow leads the penetration effect may still be observable~[\onlinecite{gorini_2006}].
We conclude by noticing that, for a nanowire in the Zeeman-dominated regime [see Fig.\ref{Fig2}(c)], the nanowire band bottom (\ref{Eminprime-Zeeman}) coincides with the lead band bottom, and only the first two cases (b) and (c) are possible.

\section{Effects of a magnetic field parallel to the Rashba field}
\label{sec-5}
So far, we have considered a magnetic field $h_x$  perpendicular  to the direction $z$ of the Rashba field. Here we wish to discuss the effects of a magnetic field component $h_z$ parallel to such direction. Again, we first analyze the effects in the bulk, and then consider the whole system nanowire+leads with a finite nanowire length. 
\begin{figure} 
\centering
\includegraphics[width=\columnwidth]{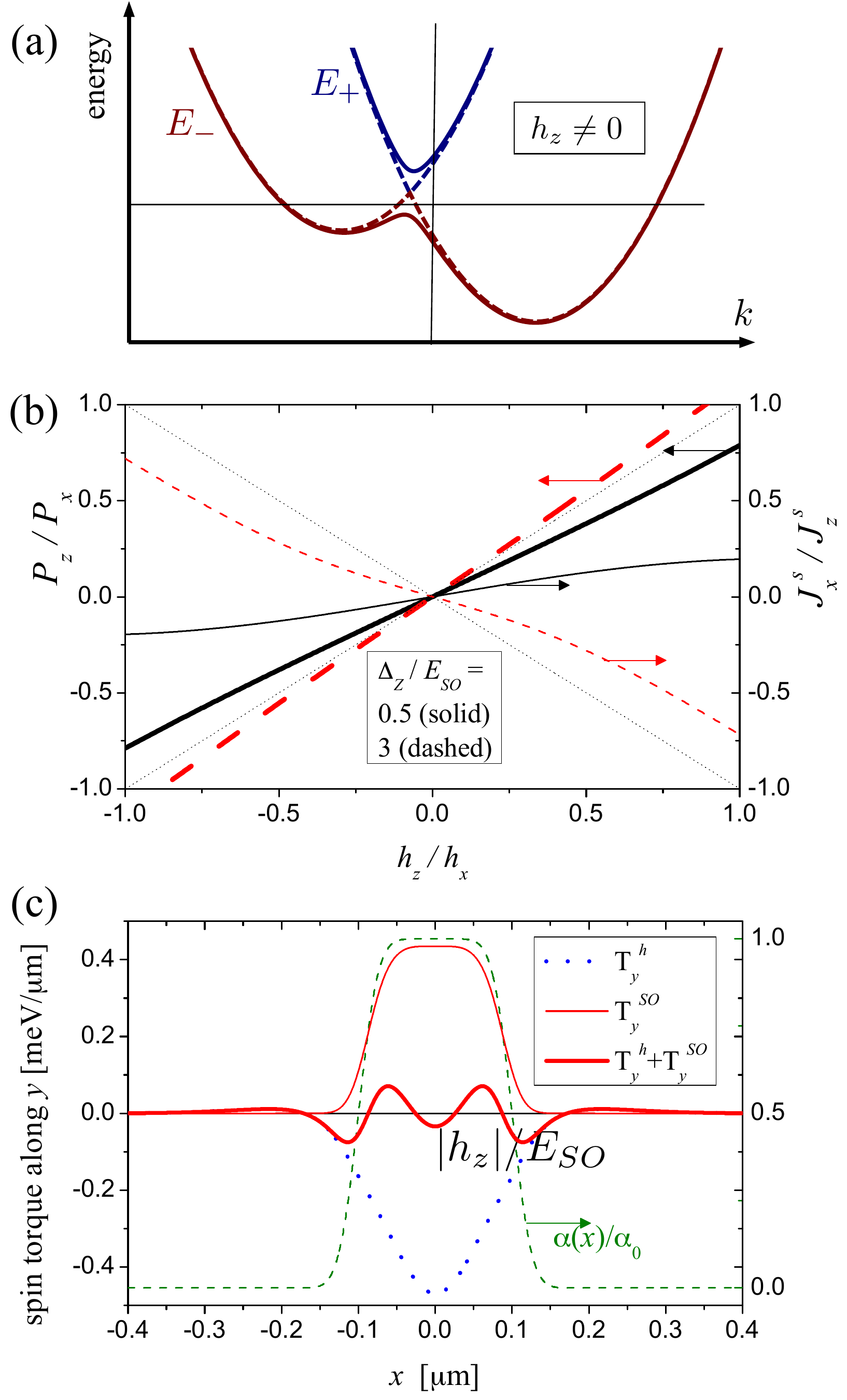}
\caption{(Color online) \label{Fig8-hpara} The effect of an additional field component ${h_z}$ parallel to the Rashba direction $z$. (a) The spectrum of the nanowire bulk is no longer symmetric in $k$ [see Fig.\ref{Fig2} for comparison].  (b)  The thick curves describe the zero temperature behavior of the bulk ratio $P_y/P_x$ as a function of $h_z/h_x$, for $\Delta_Z/E_{SO}=0.5$ (thick black solid) and $\Delta_Z/E_{SO}=3$ (thick red dashed), and show that  the spin polarization $\mathbf{P}$ is not collinear to the magnetic field $\mathbf{h}$. The reader should refer to the left vertical axis.  The thin curves describe the bulk ratio $J^s_x/J^s_z$ as a function of $h_z/h_x$, for $\Delta_Z/E_{SO}=0.5$ (thin black solid) and $\Delta_Z/E_{SO}=3$ (thin red dashed), and show that the polarization of the equilibrium spin current is not perpendicular to the applied magnetic field $\mathbf{h}$. The reader should refer to the right vertical axis. The thin dotted curves are the separatrices and serve as guides to the eye.
 (c) The space profile of the orthogonal components ${\rm T}^h_y$ of the spin torque (dotted curve), ${\rm T}^{SO}_y$ of the spin-orbit torque (thin solid curve), and  ${\rm T}^h_y+{\rm T}^{SO}_y$ the total torque (thick solide curve) in the case of a short nanowire $L_w=200\,{\rm nm}$  contacted to leads, with $E_{SO}=0.3\,  {\rm meV}$, $\Delta_Z=0.4 \,{\rm meV}$, $h_z=0.2 \,{\rm meV}$, $\mu=-0.2 \,{\rm meV}$ and $T=50\,{\rm mK}$. }
\end{figure}
\subsection{Effects of $h_z$ in the nanowire bulk} 
\subsubsection{Magnetic field  parallel to the spin-orbit field}
When  the magnetic field is applied {\it only}  along the Rashba field direction $z$ (${h_z} \neq 0$, ${h_x}=0$), the spin quantization axis remains $z$ for all electronic states, similarly to the  case of a purely Rashba coupling. 
In this case, however, the spin-$\uparrow$ and spin-$\downarrow$ bands are further shifted vertically in opposite directions by the parallel Zeeman energy ${h_z}$ [see dashed curves in Fig.\ref{Fig8-hpara}(a)]. Since the spectrum is no longer symmetric in $k$, there is an energy range where only electron states with a spin direction exist, giving rise to a net spin polarization along the direction $z$ of the magnetic field,
\begin{equation}\label{Sz-hpara}
\mathbf{P}=(0,0,P_z) \,\neq 0 \quad,
\end{equation} 
where $P_z=(\rho_\uparrow-\rho_\downarrow)/(\rho_\uparrow+\rho_\downarrow)$ and $\rho_{\uparrow,\downarrow}=(2\pi)^{-1}\!\! \int dk f^\circ(E_{\uparrow,\downarrow}(k))$ is the equilibrium spin-resolved particle density. 
In particular, at zero temperature, one obtains
\begin{equation}
\rho_{\uparrow,\downarrow}=\frac{1}{\pi} \sqrt{\frac{2 m^*(\mu+E_{SO} \pm h_z)}{\hbar^2}}\, \theta(\mu+E_{SO} \pm h_z)
\end{equation}
where $\theta$ is the Heaviside function, implying that $P_z \simeq  h_z/(4 (\mu+E_{SO}))$ for weak fields $|h_z| \ll \mu+E_{SO}$, while $P_z =\mbox{sgn}(h_z)$ for large fields $|h_z| \ge \mu+E_{SO}$.

The spin current $\mathbf{J}^s$ instead vanishes. Indeed, despite the presence of a spin polarization, at any energy $E$, each state with a definite spin orientation and group velocity has a partner state with the same spin orientation and opposite group velocity:  In each spin channel $\sigma=\uparrow,\downarrow$ the current $J_\sigma$  vanishes, and so do the equilibrium charge current  and spin current, as a straightforward evaluation of Eq.(\ref{Jsx-gen}) and Eq.(\ref{Jsz-gen}) shows.
Furthermore, substituting $\sin\theta_k=0$ and ${h_x}=0$ into Eqs.(\ref{TSOy-gen}) and (\ref{Thy-gen}), the torques are also vanishing, and one has  
\begin{equation}\label{other-hparallel}
\mathbf{J}^s={\rm\mathbf{T}}^{h}={\rm\mathbf{T}}^{SO}=0\quad.
\end{equation}
As a consequence, a magnetic field $h_z$ parallel to the Rashba spin-orbit direction cannot give rise, alone, to any equilibrium spin current or torques.

\subsubsection{Magnetic field with components parallel  and perpendicular to the spin-orbit field}
When the field component ${h_z}$ parallel to the Rashba direction is {\it added} to a  component ${h_x}$ perpendicular to the Rashba spin-orbit field, the spectrum depicted by the dashed curves of Fig.\ref{Fig8-hpara}(a)  modifies into the one described by the solid curves, where the asymmetry in $k$ appears as the essential difference from Fig.\ref{Fig2}(b). As compared to the case of purely perpendicular field $h_x$ discussed in sec.\ref{sec-perp-field}, the additional component ${h_z}$ has mainly three effects on the spin sector. 

The first one is quite expected, and is the appearance of  an additional spin polarization component along $z$,
\begin{equation}\label{spin-density-h-all}
\mathbf{P}=(P_x,0,P_z)
\end{equation}
with $P_z$ being obtained from Eqs.(\ref{P-def}), Eq.(\ref{Sz-gen}) and Eq.(\ref{charge-density-def}). Notably, because of the interplay between $h_z$ and the RSOC, the spin polarization $\mathbf{P}$ {\it is not collinear} to the magnetic field $\mathbf{h}$, as can be seen by the two thick curves in Fig.\ref{Fig8-hpara}(b), which describe the behavior of $P_z/P_x=S_z/S_x$ as a function of the ratio $h_z/h_x$ (the thin dotted separatrix lines are meant as  guides to the eye). Explicitly, the thick black solid curve refers to $\Delta_Z/E_{SO}=0.5$, while the thick red dashed curve to $\Delta_Z/E_{SO}=3$. The reader should  refer to the left vertical axis.

Secondly, $h_z$ thereby modifies the polarization of the equilibrium spin current  $\mathbf{J}^s$, since a component $J^s_x$ polarized along the nanowire axis adds up to the component $J^s_z$ along the Rashba field  caused by $h_x$,
\begin{equation}\label{spin-current-h-all}
\mathbf{J}^s=(J^s_x,0,J^s_z)\quad.
\end{equation}
In particular, when $h_z \neq 0$, the bulk polarization of $\mathbf{J}^s$ is no longer perpendicular to the applied magnetic field $\mathbf{h}$, as shown by the thin  curves in Fig.\ref{Fig8-hpara}(b) that display the ratio $J_x^s/J^s_y$ as a function of $h_z/h_x$. The thin black solid curve refers to $\Delta_Z/E_{SO}=0.5$, while the thin red dashed curve to $\Delta_Z/E_{SO}=3$. The reader should refer to the right vertical axis, with the thin dotted lines indicating the separatrix lines. 

The third effect caused by the component $h_z$ is the  emergence of torques also in the bulk of the nanowire. Indeed, since $\mathbf{S}$ is not collinear to the magnetic field $\mathbf{h}$, the spin torque  ${\rm \mathbf{T}}^{h}$ is no longer vanishing [see  Eq.(\ref{Th-def})], and exhibits a component ${T}^{h}_y$ perpendicular to the substrate plane, where $\mathbf{S}$  and $\mathbf{h}$ lie. Then, a spin-orbit torque ${\rm \mathbf{T}}^{SO}=(0,{T}^{SO}_y,0)$ must cancel such contribution, as dictated by Eq.(\ref{torque-cancellation}), i.e.
\begin{equation}\label{Th-andTSO-neq0}
(0,{T}^{h}_y,0)={\rm \mathbf{T}}^h =  - {\rm \mathbf{T}}^{SO} \,\neq 0\quad.
\end{equation}
This is thus qualitatively different from the case (\ref{torques-case3}) of a field along the nanowire axis.


\subsection{Effects of $h_z$ in the nanowire contacted to leads} 
By adopting again the inhomogeneous RSOC profile~(\ref{alpha-profile}), one can analyze the effects of $h_z$ for a nanowire contacted to the leads. In this case the  effects  of the finite length $L_w$ are particularly visible for a short nanowire, $L_w=200\,{\rm nm}$, as shown by Fig.~\ref{Fig8-hpara}(c), which shows the space profile of the torque  $y$-component,  orthogonal to substrate plane. Explicitly, the  dotted curve   describes  the behavior of the spin torque ${\rm T}^h_y$, the thin solid curve the spin-orbit torque ${\rm T}^{SO}_y$, while the thick  solid curve the total torque ${\rm T}^h_y+{\rm T}^{SO}_y$. As one can see,  ${\rm T}^h_y$ and ${\rm T}^{SO}_y$ take relatively large values with opposite sign near the middle  of the nanowire. However, differently from an infinitely long nanowire (homogeneous RSOC) where their contributions perfectly cancel [see Eq.(\ref{Th-andTSO-neq0})], a finite total torque appears because of the finite  length of the nanowire.

\section{Conclusions}
\label{sec-6}
In conclusion, in this paper we have analyzed a Rashba nanowire contacted to leads by using an inhomogeneous RSOC (see Fig.\ref{Fig-setup}), and we have determined  the equilibrium properties of the spin sector (spin density and polarization, spin current and torques) when a uniform magnetic field is applied to the nanowire. Differently from a 2DEG with RSOC, in a nanowire the mere presence of a RSOC yields a trivial behavior of the spin sector, with all spin quantities vanishing [see Eq.(\ref{case-1})]. In contrast, here we have found that the interplay between the inhomogeneous RSOC and a uniformly applied magnetic field leads to interesting effects already at equilibrium. 

Focussing first on the bulk of the nanowire  we have shown that, when the magnetic field is applied along the nanowire axis, i.e. perpendicularly to the Rashba field direction, a  spin current polarised along the Rashba field direction flows even at equilibrium. Differently from the equilibrium spin current found in nanorings with magnetic field texture[\onlinecite{loss-goldbart_PRL_1990},\onlinecite{loss-goldbart_PRB_1992}], this is  a bulk effect that does not disappear in the limit of infinitely long wire. It exists only if {\it both} RSOC and magnetic field are present, since under these conditions a non-trivial quantum correlation between the velocity direction and the spin orientation  appears, so that in the system ground state   electrons with opposite spin counter-propagate. Remarkably,
the  analogy between the nanowire states inside the magnetic gap and the helical states of a QSH system, which is useful  to intuitively understand why such states  give rise to an equilibrium spin current   [see Fig.\ref{Fig2}(b)], is not sufficient to explain the actual behavior of the equilibrium spin current  (see Fig.\ref{Fig3-Jsz-hperp}), which persists far beyond   the Rashba-dominated regime where such analogy holds.  Indeed we have shown that also the states outside the magnetic gap provide an unexpectedly significant contribution to the equilibrium spin current,  due to  the lack of factorization between velocity and spin [see Eq.(\ref{inequality})] that characterize the Rashba nanowire eigenstates under a magnetic field. Such equilibrium spin current is tunable by both the applied magnetic field (see Fig.\ref{Fig3-Jsz-hperp}) and the chemical potential (see Fig.\ref{Fig4-Jsz-mu}).

Then, considering the whole system nanowire+lead with the inhomogeneous RSOC profile, we find that  inside the nanowire the equilibrium spin current exhibits weak oscillations around the finite bulk value,
characterized by a wavelength (\ref{lambda-osc-def}), while $\mathbf{J}^s$ dies out in the bulk of the leads, where RSOC eventually vanishes.
Interesting effects emerge at the nanowire/lead interfaces, namely the appearance of a spin polarization component $P_y$ {\it orthogonal} to both the magnetic field and the RSOC field, and a localized spin torque $T^h_z$ related to it  (see Fig.\ref{Fig-inhomo-long} for a long nanowire and Fig.\ref{Fig-inhomo-short} for a short nanowire).  These interface spin torques act as sources and sinks of the  spin current, which is carried in the nanowire by spin bound states  sketched in Fig.\ref{Fig-setup}(b), while the leads play the role of magnetic barriers. 
The inhomogeneity of the RSOC thus leads to effectively spin-active interfaces, even under a uniform  magnetic field and in the absence of magnetic material. The appearance of this orthogonal spin polarization at the nanowire edges also impacts in the interpretation of results concerning proximized nanowires where Majorana fermions should be observable, as we shall discuss below. 
Notably, the orthogonal spin polarization and the spin torque partially penetrate into the leads for a lengthscale  $\lambda_{pen}$ that depends on the three possible configurations of chemical potential  (see Fig.\ref{Fig7-interface}). 

Finally, we have analyzed the effect of a magnetic field  not collinear with the nanowire axis. A magnetic field  $h_z$ directed along the Rashba field leads to essentially trivial results, as the only non-vanishing quantity is a spin polarization component $P_z$ [see Eq.(\ref{other-hparallel})]. However, when  $h_z$ is added to a component $h_x$, different effects emerge in the nanowire bulk: the spin polarization $\mathbf{P}$ becomes not collinear with the magnetic field,  the polarization of the equilibrium spin current $\mathbf{J}^s$ is no longer orthogonal to the magnetic field  [see Fig.\ref{Fig8-hpara}(b)], and both a spin torque and a spin-orbit torque appear. While the contributions of the two torques perfectly cancel for a homogeneous infinitely long nanowire, for a realistic nanowire with finite length a residual of torque survives [see Fig.\ref{Fig8-hpara}(c)].

\subsection{Experimental realizations} The setup of Fig.\ref{Fig-setup}  is currently realized  with   e.g.  InSb nanowires or InAs nanowires, and  realistic values for the involved   parameters have been used here for the plots. In the  case of InSb, for instance, the effective mass and the  $g$-factor are  $m^* \simeq 0.015 m_e$ and $g\simeq 50$, respectively, while the value of the RSOC depends on the specific implementation and experimental conditions and  can be widely tunable, e.g. $\alpha \sim (0.03 \div 1)\, {\rm eV} \, {\rm \AA}$~[\onlinecite{wimmer_2015},\onlinecite{kouwenhoven_2012},\onlinecite{nilsson_2009,xu_2012,kouwenhoven_PRL_2012}].
In the case of InAs nanowires $m^*\simeq 0.022\,m_e$, $g\simeq 20$ and $\alpha \sim (0.05 \div 0.3) \, {\rm eV \, \AA}$~ [\onlinecite{gao-2012,heiblum_2012,ensslin_2010,joyce_2013}]. The spin-orbit energy $E_{SO}$ resulting from these values  [see Eq.(\ref{ESO-def})] and used in the plots is some fractions of ${\rm meV}$. We have taken similar values  for the Zeeman gap energy~$\Delta_Z$, which corresponds to a magnetic field of  some hundreds of ${\rm mT}$, while the temperature of $50\,{\rm mK}$ is taken from  recent low temperature experiments~[\onlinecite{kouwenhoven_2012},\onlinecite{heiblum_2012}].

As far as the measurements of the predicted spin polarization is concerned, spatially resolved  detection of spin orientation with nanometer scale resolution can be performed with various methods such as magnetic resonance force microscopy~[\onlinecite{chui_2004},\onlinecite{hammel_2015}],   spin-polarized scanning electron microscopy~[\onlinecite{koike_1985},\onlinecite{kohashi_2015}], by exploiting quantum dots as probes~[\onlinecite{katsumoto_2009},\onlinecite{tarucha_2012}], or also electrically by potentiometric measurements using ferromagnetic detector contacts~[\onlinecite{jonker_2014},\onlinecite{wang_2014}]. 

Concerning spin currents, they are customarily measured by exploiting the Kerr effect [\onlinecite{awshalom_science_2004},\onlinecite{awshalom_naturephys_2005}], the inverse spin Hall effect [\onlinecite{tinkham_2006}] or the charge imbalance voltage appearing on the normal lead~[\onlinecite{zutic_2011}]. Also, pure spin current  can be indirectly revealed via  or by optical detection exploiting a polarized light beam~[\onlinecite{zhu_2010}]. 
However, for the detection of   spin current at equilibrium, these methods are not straightforwardly applicable, and various alternative principles have been proposed. Equilibrium spin current can for instance be detected by measuring the thereby generated spin torque[\onlinecite{chen_PRB_2006}].
Another proposal is based on detection of the mechanical deformations that the equilibrium spin current induces onto  the substrate  underneath the Rashba spin-orbit material: A mechanical cantilever magnetometer with an integrated 2D electron system can be used to this purpose[\onlinecite{sonin_PRL_2007}]. Other works suggested   that an equilibrium spin current  can be measured through the electric field  it produces, similarly to the case of a charge current that generates a magnetic field [\onlinecite{wang_PRB_2008},\onlinecite{loss-meier_2004}]. 
These  methods can in principle apply to the nanowire as well. In particular, as emphasized in Sec.\ref{sec-4-B}, the spin torque generated at the interfaces [see  Figs.\ref{Fig-inhomo-long}(b) and \ref{Fig-inhomo-short}(b)] is precisely the hallmark of the equilibrium spin current present in the nanowire bulk.
It is thus plausible that the predicted spin properties are at experimental reach.\\

\subsection{Future developments} We  conclude by outlining some possible future developments of the present work. 
We first notice that, although  we have focused here on an inhomogeneous RSOC profile that vanishes in the leads, it is straightforward to generalize our results to more complex hybrid structures, such as interfaces between regions where RSOC takes different signs, which may possibly be obtained by coupling the nanowire with different finger gates.  

Secondly, the present analysis of the equilibrium properties represents the first  necessary  step for any out of equilibrium analysis, such as the effects of a time-dependent RSOC induced by ac voltages applied to the gates. In particular, in order to correctly determine the impact of the external drive  on the spin current,  the equilibrium contribution found here must be subtracted, for otherwise one could mistakenly interpret a non-vanishing spin current as entirely due to the out of equilibrium conditions[\onlinecite{rashba_2003}].  In particular, the value of the equilibrium spin current also identifies the limit of sensitivity for the out of equilibrium spin current contribution.

Finally, a promising future development of this work would be the inclusion of superconducting (SC) leads. For a nanowire with homogeneous RSOC  it is well known that, in appropriate   parameter ranges, the interplay bewteen an $s$-wave superconducting coupling and RSOC leads the proximized nanowire to effectively behave as a $p$-wave topological superconductor, hosting Majorana fermions at the boundaries with normal regions\cite{}. 
In this respect, there are two aspects that we would like to comment about. The first one is related to the spin polarization.  
For a proximized nanowire with homogeneous RSOC, the behavior of spin polarization has been shown to provide useful insights about the topological transition, and it has been recently suggested that the spin polarization of Majorana fermions could be exploited to distinguish them from ordinary fermion states of the nanowire[\onlinecite{bena-simon_2012,black-schaffer_2015,loss_PRB_2017,domanski_scirep_2017}]. Let us compare this case to the spin polarization shown in Fig.\ref{Fig-inhomo-long}(a) that we find for a normal nanowire without proximity effect and with inhomogeneous RSOC. In both cases a spin polarization orthogonal to the nanowire axis arises at the ends of the nanowire, where it takes opposite values. However, while in the former case the spin polarization is an effect of the topological phase, in the latter case the nanowire is   purely due to inhomogeneity of the RSOC. Such comparison  shows that a caution should thus be taken in interpreting experimental analysis of orthogonal spin polarization at the boundaries as a signature of Majorana fermions.
The second aspect that we would like to point out is the interplay between
the inhomogeneus RSOC  and the superconducting coupling. Although so far most works have focused on the effect of inhomogeneities of either the scalar potential or the superconducting parameter [\onlinecite{romito_PRL_2011,romito_PRB_2011,sen_2013,dassarma_2015,loss_PRB_2016,domansky_PRB_2017,domansky-ptok_PRB_2017}], we expect that inhomogeneities of the RSOC can impact as well, in a twofold way.
In the first instance, the very presence of a superconducting film deposited on top  of a nanowire portion can produce a structural inversion asymmetry,  altering locally the magnitude of the RSOC. Such inhomogeneity can thus affect   the stability of the topological phase, as the topological gap crucially depends on the value of $\alpha$  in the proximized nanowire portion[\onlinecite{alicea_PRB_2010}]. Furthermore, when the RSOC profile changes sign, either because of the very presence of the  SC film or because of locally applied finger gate voltages,  fermionic localized states form at the sign change region. It has been recently shown[\onlinecite{loss_EPJB_2015}] that, under suitable conditions,  these states have zero energy and can lead to an hybridization of the otherwise spatially separated Majorana fermions. This effect would  split the Majorana off from zero energy and disguise, or even jeopardize, their observability.  For these reasons a detailed investigation of a proximized nanowire with inhomogeneous RSOC seems a promising research topic. Work is in progress along these lines.

\acknowledgments

Fruitful and inspiring discussions with Ch. Fleckenstein, C. Gorini, B. Trauzettel and N. Traverso are greatly acknowledged.\\

\appendix
\section{Spin equation of motion and torques}
Here we provide the proof of the spin Equation of motion Eq.(\ref{spin-cont-eq}). From the Heisenberg equation for the electron field spinor 
\begin{eqnarray}
 \partial_t \hat{\Psi}&=& \frac{1}{i \hbar}\left[-\frac{\hbar^2}{2 m^*} \partial_x^2 \hat{\Psi}-
 \frac{\sigma_z}{2 \hbar}  \left\{ \alpha(x) , \hat{p}_x  \right\} \,    \hat{\Psi}-\mathbf{h}\cdot  \boldsymbol\sigma \, \hat{\Psi}\right]  \hspace{0.5cm}
\end{eqnarray}
and the one for its adjoint $\hat{\Psi}^\dagger$, the evolution of the spin density is given by
\begin{eqnarray}
\partial_t \hat{\mathbf{S}}  
&=& \frac{\hbar}{2}  \left[ -  \frac{i\hbar}{2 m^*} \left( \partial_x^2 \hat{\Psi}^\dagger \boldsymbol\sigma\, \hat{\Psi}^{}   -\hat{\Psi}^\dagger \, \boldsymbol\sigma\,  \partial_x^2\hat{\Psi} \right) +\right.\nonumber \\
& &  \hspace{0.7cm} +    
 \frac{i}{2 \hbar^2} \hat{\Psi}^\dagger\boldsymbol\sigma  \sigma_z \left\{ \alpha(x) , \hat{p}_x  \right\} \,    \hat{\Psi} +\nonumber \\
& & \hspace{0.7cm} + \frac{i}{2 \hbar^2}   \left\{ \alpha(x) , \hat{p}_x  \right\} \hat{\Psi}^\dagger \sigma_z \boldsymbol\sigma \, \hat{\Psi}  \nonumber \\
& & \hspace{0.7cm} + \frac{i}{\hbar} \left.   \hat{\Psi}^\dagger \, \,    \left[   \boldsymbol\sigma \, , \, \mathbf{h}\cdot  \boldsymbol\sigma \right] \hat{\Psi}   \right] 
\end{eqnarray}
Rewriting
$\boldsymbol\sigma \sigma_z  =\left( \left\{\boldsymbol\sigma , \sigma_z\right\}+ \left[\boldsymbol\sigma, \sigma_z   \right]\right)/2$ and $\sigma_z \boldsymbol\sigma = \left( \{\boldsymbol\sigma , \sigma_z\} - \left[\boldsymbol\sigma, \sigma_z   \right] \right)/2$,
one obtains
\begin{eqnarray}
\partial_t \hat{\mathbf{S}} &=&\frac{\hbar}{2}  \left[      \frac{i\hbar}{2 m^*}   \partial_x \left( \hat{\Psi}^\dagger \, \boldsymbol\sigma\,  \partial_x \hat{\Psi} - \partial_x  \hat{\Psi}^\dagger \boldsymbol\sigma\, \hat{\Psi}^{}   \right)  +\right.\nonumber \\
& & \hspace{0.7cm} +    
 \frac{i}{2 \hbar^2} \hat{\Psi}^\dagger \frac{\{\boldsymbol\sigma , \sigma_z\}}{2} \left\{ \alpha(x) , \hat{p}_x  \right\}     \hat{\Psi} \,\,+ {\rm H.c.} +\nonumber \\
& &  \hspace{0.7cm} +    
 \frac{i}{2 \hbar^2} \hat{\Psi}^\dagger \frac{\left[\boldsymbol\sigma, \sigma_z   \right]}{2} \left\{ \alpha(x) , \hat{p}_x  \right\}     \hat{\Psi} \,+ {\rm H.c.}  \,\,\nonumber \\
& & \hspace{0.7cm} + \frac{i}{\hbar} \left.   \hat{\Psi}^\dagger \, \,    \left[   \boldsymbol\sigma \, , \, \mathbf{h}\cdot  \boldsymbol\sigma \right] \hat{\Psi}   \right] \label{dtS-pre}
\end{eqnarray}
A straightforward calculation leads to show that the second line in Eq.(\ref{dtS-pre}) can be rewritten  as 
\begin{eqnarray}
\lefteqn{ \frac{i}{2 \hbar^2} \hat{\Psi}^\dagger \frac{\{\boldsymbol\sigma , \sigma_z\}}{2} \left\{ \alpha(x) , \hat{p}_x  \right\}     \hat{\Psi} \,\,+ {\rm H.c.}=} & & \nonumber \\
&=&  \partial_x \left(  \frac{\alpha(x)}{\hbar} \, \hat{\Psi}^\dagger  \frac{\{\boldsymbol\sigma , \sigma_z\}}{2} \, \hat{\Psi}   \right)\label{second-line}
\end{eqnarray}
while the third line in (\ref{dtS-pre}) can be rewritten as
\begin{eqnarray}
\lefteqn{ \frac{i}{2 \hbar^2} \hat{\Psi}^\dagger \frac{\{\boldsymbol\sigma , \sigma_z\}}{2} \left\{ \alpha(x) , \hat{p}_x  \right\}     \hat{\Psi} \,\,+ {\rm H.c.}  = } & & \nonumber \\
&=& \frac{\alpha(x)}{2 \hbar} \left( \hat{\Psi}^\dagger \left[\boldsymbol\sigma, \sigma_z   \right]  \partial_x \hat{\Psi} -  \partial_x \hat{\Psi}^\dagger \left[\boldsymbol\sigma, \sigma_z   \right]  \hat{\Psi}  \right) = \nonumber \\
&=& \frac{i}{\hbar} \hat{\Psi}^\dagger \frac{\left[\boldsymbol\sigma, \boldsymbol\sigma\cdot \mathbf{h}^{SO}   \right]}{2}     \hat{\Psi} \,\, + {\rm H.c.} \quad,\label{third-line}
\end{eqnarray}
where we have used  the definition (\ref{SO-field-op}) of the Rashba field operator. 
Inserting  Eqs.(\ref{second-line}) and (\ref{third-line}) into Eq.(\ref{dtS-pre}), exploiting   the properties of Pauli matrices
$
 \left[ \boldsymbol\sigma \, , \, \mathbf{a} \cdot  \boldsymbol\sigma \right] =-2i \, \boldsymbol\sigma \, \times  \, \mathbf{a} 
$, and recalling the definitions of spin current density  (\ref{spin-current-def}), spin torque~(\ref{Th-def}) and spin-orbit torque~(\ref{TSO-def}), 
one obtains Eq.(\ref{spin-cont-eq}).


\begin{thebibliography}{99}

 

\bibitem{nitta_2015} A. Manchon, H. C. Koo, J. Nitta, S. M. Frolov, and R. A. Duine, Nature Mater. {\bf 14}, 871 (2015).


\bibitem{kane-mele2005a} C. L. Kane, E. J. Mele, Phys. Rev. Lett. {\bf 95}, 146802 (2005).
\bibitem{kane-mele2005b} C. L. Kane, E. J. Mele, Phys. Rev. Lett. {\bf 95}, 226801 (2005).
\bibitem{bernevig_science_2006} B. A. Bernevig, T. L. Hughes, and S.-C. Zhang, Science {\bf 314}, 1757 (2006).

\bibitem{vonoppen_2010} Y. Oreg,  G. Refael,  and F. von Oppen, Phys. Rev. Lett. {\bf 105}, 177002 (2010).
\bibitem{dassarma_2010} R. M. Lutchyn, J. D. Sau, and S. Das Sarma, Phys. Rev. Lett. {\bf 105}, 077001 (2010).


\bibitem{gao-2012} D. Liang and X. P.A. Gao, Nanolett. {\bf 12}, 3263 (2012).
\bibitem{wimmer_2015} I. van Weperen, B. Tarasinski, D. Eeltink, V. S. Pribiag, S. R. Plissard, E. P. A. M. Bakkers,  L. P. Kouwenhoven,  and M. Wimmer, Phys. Rev. B {\bf 91},   201413(R) (2015).
\bibitem{nygaard_2016} Z. Scher\"ubl,  G. F\"ul\"op, M. H. Madsen, J. Nyg{\aa}rd, and S. Csonka, Phys. Rev. B {\bf 94}, 035444 (2016)

\bibitem{sasaki_2017} K. Takase, Y. Ashikawa, G. Zhang, K. Tateno, and S. Sasaki, Sci. Rep. {\bf 7}, 930 (2017).
\bibitem{loss_2017} Ch. Kloeffel,  M. J. Ran\v{c}i\'{c}, and D. Loss, Phys. Rev. B {\bf 97}, 235422 (2018).

\bibitem{kouwenhoven_2012} V. Mourik,  K. Zuo,  S. M. Frolov,  S. R. Plissard,  E. P. A. M. Bakkers, L. P. Kouwenhoven, Science {\bf 336}, 1003 (2012).
\bibitem{liu_2012} L. P. Rokhinson, X. Liu, and J. K. Furdyna, Nature Phys. {\bf 8}, 795 (2012).
\bibitem{heiblum_2012} A. Das, Y. Ronen, Y. Most, Y. Oreg, M. Heiblum, and H. Shtrikman, Nature Phys. {\bf 8}, 887 (2012).
\bibitem{xu_2012} M. T. Deng,  C. L. Yu,  G. Y. Huang,  M. Larsson, P. Caroff,  and H. Q. Xu, Nanolett. {\bf 12}, 6414 (2012).
\bibitem{defranceschi_2014} E. J. H. Lee, X. Jiang, M. Houzet, R. Aguado, C. M. Lieber, and S. De Franceschi, Nature Nanotech. {\bf 9}, 79 (2014).
\bibitem{marcus_2016} S. M. Albrecht, A. P. Higginbotham, M. Madsen, F. Kuemmeth, T. S. Jespersen, J. Nyg{\aa}rd, P. Krogstrup, and C. M. Marcus, Nature {\bf 531}, 206 (2016).
\bibitem{marcus_science_2016} M. T. Deng, S. Vaitiekenas,  E. B. Hansen,  J. Danon,  M. Leijnse,  K. Flensberg,  J. Nyg{\aa}rd,  P. Krogstrup, and  C. M. Marcus, Science {\bf 354}, 1557 (2016).


\bibitem{tokatly_PRB_2017} J. Borge and I. V. Tokatly, Phys. Rev. B {\bf 96}, 115445 (2017).


\bibitem{brataas_2007} A. Brataas, A G Mal'shukov, and Y. Tserkovnyak, New J. Phys. {\bf 9}, 345 (2007).
\bibitem{bercioux_2012} D. Bercioux, D. F. Urban, F. Romeo, and R. Citro, Appl. Phys. Lett. {\bf 101}, 122405 (2012).

\bibitem{raimondi_2014} G. Seibold, S. Caprara, M. Grilli, and R. Raimondi, J. Magn. Magn. Mater.  {\bf 440}, 63 (2017).
\bibitem{raimondi_2017} N. Bovenzi, S. Caprara, M. Grilli, R. Raimondi, N. Scopigno, G. Seibold, pre-print ArXiv:1704.01852
\bibitem{sherman_2018} S. Kud{\l}a, A. Dyrda{\l}, V. K. Dugaev, E. Ya. Sherman, and J. Barna\'{s}, Phys. Rev. B {\bf 97}, 245307 (2018).
 

\bibitem{sanchez_2006} D. S\'anchez, L. Serra, Phys. Rev. B {\bf 74} 153313 (2006).
\bibitem{sanchez_2008} D. S\'anchez, L. Serra,  and M.-S. Choi, Phys. Rev. B {\bf 77},    035315 (2008).
\bibitem{sherman_2011} M. M. Glazov, and E. Ya. Sherman, Phys. Rev. Lett. {\bf 107}, 156602 (2011).
\bibitem{sherman_2013} A. F. Sadreev, and E. Y. Sherman, Phys. Rev. B {\bf 88}, 115302 (2013).
\bibitem{sherman_2017} M. Modugno, E. Ya. Sherman, and V. V. Konotop, Phys. Rev. A   {\bf 95}, 063620 (2017).
\bibitem{loss_EPJB_2015} J. Klinovaja, and D. Loss, Eur. Phys. J. {\bf B 88}, 62 (2015)


\bibitem{rashba_2003} E. I. Rashba,  Phys. Rev. B {\bf 68} 241315(R) (2003). 
\bibitem{governale_2003} J. Splettstoesser, M. Governale, and U. Z\"ulicke, Phys. Rev. B {\bf 68}, 165341 (2003).
\bibitem{balseiro_2005} G. Usaj and C. A. Balseiro, Europhys. Lett. {\bf 72}, 631 (2005).
\bibitem{sonin_PRB_2007} E. B. Sonin,  Phys. Rev. B {\bf 76}, 033306 (2007).
\bibitem{sonin_PRL_2007} E. B. Sonin,  Phys. Rev. Lett. {\bf 99}, 266602 (2007).
\bibitem{wang_PRL_2007} Q.-F. Sun, X. C. Xie, J. Wang, Phys. Rev. Lett. {\bf 98}, 196801 (2007).
\bibitem{wang_PRB_2008} Q.-F. Sun, X. C. Xie, J. Wang, Phys. Rev. B {\bf 77}, 035327 (2008).
\bibitem{tokatly_2008} I. V. Tokatly, Phys. Rev. Lett. {\bf 101}, 106601 (2008).
\bibitem{sablikov_2008} V. A. Sablikov, A. A. Sukhanov, and Y. Ya. Tkach, Phys. Rev. B {\bf 78}, 153302 (2008).
\bibitem{liang_PLA_2008} F. Liang Y. G. Shen, Y. H. Yang, Phys. Lett. A {\bf 372}, 4634 (2008).
\bibitem{medina_2010} B. Berche, C. Chatelain, E. Medina, Eur. J. Phys. {\bf 31}, 1267 (2010).
\bibitem{sonin_2010} E. B. Sonin, Adv. Phys. {\bf 59}, 181 (2010).
\bibitem{nakhmedov_2012} E. Nakhmedov, and O. Alekperov, Phys. Rev. B {\bf 85}, 153302 (2012).
\bibitem{liu_2014} H. Zhang, Z. Ma, and J.F. Liu, Sci. Rep. {\bf 4}, 6464 (2014).
\bibitem{liang_PLA_2015} F. Liang, B.-L. Gao, G. Hu, Y. Gu, and N. Xu, Phys. Lett. A {\bf  379}, 3114  (2015).


\bibitem{malshukov_2003} A. G. Mal'shukov, C. S. Tang, C. S. Chu, and K. A. Chao, Phys. Rev. B {\bf 68}, 233307 (2003).
\bibitem{wang_2006} J. Wang, and K. S. Chan, Phys. Rev. B {\bf 74}, 035342 (2006).


\bibitem{ojanen_2012} T. Ojanen, Phys. Rev. Lett. {\bf 109}, 226804 (2012)
\bibitem{rainis-loss_PRL_2014} D. Rainis, A. Saha, J. Klinovaja, L. Trifunovic,  and D. Loss, Phys. Rev. Lett. {\bf 112}, 196803 (2014).
\bibitem{rainis-loss_PRB_2014} D. Rainis and D. Loss, Phys. Rev. B {\bf 90}, 235415 (2014).
\bibitem{aguado_2015} J. Cayao, E. Prada,  P. San-Jose,  and R. Aguado, Phys. Rev. B {\bf 91}, 024514 (2015).

\bibitem{malshukov_2005} C. S. Tang,  A. G. Mal'shukov,  and K. A. Chao, Phys. Rev. B {\bf 71}, 195314 (2005).

\bibitem{loss_2016} J. Klinovaja, P. Stano, and D. Loss, Phys. Rev. Lett. {\bf 116}, 176401 (2016).




\bibitem{bena-simon_2012} D. Sticlet, C. Bena, and P. Simon, Phys. Rev. Lett. {\bf 108}, 096802 (2012).
\bibitem{nota-bulk} The expression ``bulk" is meant here in the 1D sense of the nanowire axis nanowire direction.
 
 
 
\bibitem{loss-goldbart_PRL_1990} D. Loss,  P. M. Goldbart, A. V. Balatsky, Phys. Rev. Lett {\bf 65}, 1655 (1990).
\bibitem{loss-goldbart_PRB_1992} D. Loss, and P. M. Goldbart, Phys. Rev. B {\bf 45}, 13544 (1992).
\bibitem{aronov_1993} A. G. Aronov, and Y. B. Lyanda-Geller, Phys. Rev. Lett. {\bf 70}, 343 (1993).
\bibitem{hugues_2014} Q. Meng, S. Vishveshwara, and T. L. Hughes, Phys. Rev. B {\bf 90},  205403 (2014).

\bibitem{alicea_PRB_2010} J. Alicea, Phys. Rev. B {\bf 81}, 125318 (2010).

\bibitem{nota-penetration} In the regime (a) of chemical potential the penetration length for the  spin density $S_y$ reads $\lambda^{S_y}_{pen} \simeq  \pi\hbar/ \left(\sqrt{2m^*(-\mu +\Delta_Z)} +\sqrt{2m^*(-\mu -\Delta_Z)}\right)$, and is shorter than the one for the spin polarization (\ref{lambdaprox-a}).




\bibitem{fert_2001} A. Fert and H. Jaffr\`{e}s, Phys. Rev. B {\bf 64}, 184420 (2001).
\bibitem{gorini_2006} P. Schwab, M. Dzierzawa,  C. Gorini, and R. Raimondi, Phys. Rev. B {\bf 74}, 155316  (2006).
%
\bibitem{nilsson_2009} H. A. Nilsson, Ph. Caroff, C. Thelander, M. Larsson, J. B. Wagner,  L.-E. Wernersson, L. Samuelson,  and H. Q. Xu, Nanolett. {\bf 9}, 3151 (2009).
\bibitem{kouwenhoven_PRL_2012} S. Nadj-Perge,  V. S. Pribiag,  J.W. G. van den Berg, K. Zuo, S. R. Plissard,  E. P. A. M. Bakkers, S. M. Frolov,  and L. P. Kouwenhoven, Phys. Rev. Lett. {\bf 108}, 166801 (2012).
 
 
\bibitem{ensslin_2010} P. Roulleau, T. Choi, S. Riedi, T. Heinzel, I. Shorubalko, T. Ihn, and K. Ensslin, Phys. Rev. B {\bf 81}, 155449 (2010) 

\bibitem{joyce_2013} H. J. Joyce, C. J. Docherty, Q. Gao, H. H. Tan, C. Jagadish, J. Lloyd-Hughes, L. M. Herz and M. B. Johnston, Nanotechnology {\bf 24}, 214006 (2013).

 


\bibitem{chui_2004} D. Rugar, R. Budakian, H. J. Mamin and B. W. Chui, Nature {\bf 430}, 329 (2004).
\bibitem{hammel_2015} J. Cardellino, N. Scozzaro, M. Herman, A  J. Berger, C. Zhang, K. C. Fong, C. Jayaprakash, D. V. Pelekhov and P. C. Hammel, Nature Nanotech. {\bf 9}, 343 (2015).
\bibitem{koike_1985} K. Koyke, H. Matsuyama, H. Todokoro, and K. Hayakawa, Jpn. J. Appl. Phys. {\bf 24}, 1078 (1985).
\bibitem{kohashi_2015} T. Kohashi, J. Magn. Soc. Jpn., {\bf 39}, 131  (2015).

 \bibitem{katsumoto_2009} T. Otsuka, E. Abe, Y. Iye, and S. Katsumoto, Phys. Rev. B {\bf 79}, 195313 (2009).
 \bibitem{tarucha_2012} T. Otsuka,  Y. Sugihara, J. Yoneda, S. Katsumoto, and S. Tarucha, Phys. Rev B {\bf 86}, 081308 (2012).

 \bibitem{jonker_2014}  C. H. Li, O. M. J. van 't Erve, J. T. Robinson, Y. Liu, L. Li, and  B. T. Jonker, Nature Nanotech. {\bf 9}, 218 (2014).
 \bibitem{wang_2014} J.  Tang, L.-T.  Chang, X.  Kou, K. Murata, E. S.  Choi, M. Lang, Y. Fan, Y. Jiang, M. Montazeri, W. Jiang, Y. Wang, L. He, and  K. L.  Wang, Nanolett. {\bf 14}, 5423 (2014).



\bibitem{awshalom_science_2004} Y. K. Kato, R. C. Myers, A. C. Gossard, and D. D. Awschalom, Science {\bf 306}, 1910 (2004)
\bibitem{awshalom_naturephys_2005} V. Sih, R. C. Myers, Y. K. Kato, W. H. Lau, A. C. Gossard, and D. D. Awschalom, Nat. Phys. {\bf 1}, 31 (2005).
\bibitem{tinkham_2006} S. O. Valenzuela and M. Tinkham, Nature {\bf 442}, 176 (2006).

\bibitem{zutic_2011} I. \v{Z}uti\'{c} and H. Dery, Nature Mater. {\bf 10}, 647 (2011).
\bibitem{zhu_2010} J. Wang, R.-B. Liu, B.-F. Zhu, J. Supercond. Nov. Magn. {\bf 23}, 53  (2010).

 
\bibitem{chen_PRB_2006} T.-W. Chen,  C.-M. Huang,  and G. Y. Guo, Phys. Rev. B {\bf 73}, 235309 (2006).
\bibitem{loss-meier_2004} F. Meier, and D. Loss, Phys. Rev. Lett. {\bf 90}, 167204 (2003).


\bibitem{black-schaffer_2015} K. Bj\"ornson, S. S. Pershoguba,  A. V. Balatsky, and A. M. Black-Schaffer, Phys. Rev. B {\bf 92}, 214501 (2015).
\bibitem{loss_PRB_2017}  P. Szumniak, D. Chevallier, D. Loss,  and J. Klinovaja, Phys. Rev. B {\bf 96}, 041401(R) (2017).
\bibitem{domanski_scirep_2017}  M. M. Ma\'{s}ka, and T. Doma\'{n}ski, Sci. Rep. {\bf 7}, 16193 (2017).
\bibitem{romito_PRL_2011}  P. W. Brouwer, M. Duckheim, A. Romito, and F. von Oppen, Phys. Rev. Lett. {\bf 107}, 196804 (2011). 
\bibitem{romito_PRB_2011}   P. W. Brouwer, M. Duckheim, A. Romito, and F. von Oppen, Phys. Rev. B {\bf 84}, 144526 (2011). 
\bibitem{sen_2013}  W. DeGottardi,  D. Sen,  and S. Vishveshwara, Phys. Rev. Lett. {\bf 110}, 146404 (2013). 
\bibitem{dassarma_2015}  H.-Y. Hui, J. D. Sau, and S. Das Sarma, Phys. Rev. B {\bf 92}, 174512 (2015).
\bibitem{loss_PRB_2016}   S. Hoffman, J. Klinovaja, and D. Loss, Phys. Rev. B {\bf 93}, 165418 (2016). 
\bibitem{domansky_PRB_2017}  M. M. Ma\'{s}ka,  A. Gorczyca-Goraj,  J. Tworzyd{\l}o,  and   T. Doma\'{n}ski, Phys. Rev. B {\bf 95}, 045429 (2017).
\bibitem{domansky-ptok_PRB_2017} A. Ptok, A. Kobia{\l}ka,   and T. Doma\'{n}ski, Phys. Rev. B {\bf 96}, 195430 (2017). 

\end{thebibliography}
\end{document}